\documentclass[a4paper,prd,superscriptaddress,11pt,noshowkeys,notitlepage,nofootinbib]{revtex4}

\usepackage[english]{babel}
\usepackage{mathtools}
\usepackage[]{units}
\usepackage[top=2cm,bottom=2cm,left=3cm,right=3cm,marginparwidth=1.75cm]{geometry}
\usepackage[usenames]{xcolor}
\usepackage{amsfonts,amsmath,amssymb}

\usepackage{graphicx}
\usepackage[colorlinks=true, allcolors=blue]{hyperref}
\usepackage{subcaption}

\newcommand{\amu}{\ensuremath{a_\mu}}

\newcommand{\amuBSM}{\ensuremath{\amu^{\text{BSM}}}}

\begin{document}

\title{On the viability of a light scalar spectrum for 3-3-1 models}

\author{A.~L.~Cherchiglia}
\thanks{E-mail: alche@unicamp.br}
\author{O.~L.~G.~Peres}
\thanks{E-mail: orlando@ifi.unicamp.br}
\affiliation{Instituto de Física Gleb Wataghin, Universidade Estadual de Campinas,
Rua Sérgio Buarque de Holanda, 777, Campinas, SP, Brasil}

\begin{abstract}

In this work we study an effective version of the 3-3-1 model, in which the particle content is the same of the 2HDM. We show that the inherited structure from the $SU(3)_C \otimes SU(3)_L \otimes U(1)_X$ gauge group has a series of consequences, the most relevant one being the prediction of the masses of the neutral scalar to be of the order or lower than the mass of the charged scalar. Given current constraints from collider searches, B-physics, as well as theoretical constraints such as perturbativity of quartic couplings and stability of the scalar potential, we find that the new scalars cannot be lighter than 350 GeV.

\end{abstract}

\maketitle

\section{Introduction}

In the present, the quest for Beyond Standard Model (BSM) signals is as strong as ever. From a collider perspective, the LHC has just resumed its operations aiming to increase the amount of data in a variety of processes, with the hope to pinpoint deviations from the SM predictions. Although the experimental program is the one responsible to unveil the path towards BSM, it is of utmost importance to scrutinize present theoretical proposals aiming to identify the kind of signals that they can (or not) address. Among these models, an extension of the SM gauge group is compelling since it may address some of the open problems of the SM as well as provide a rich phenomenology testable at the LHC. In particular, we will consider in this contribution the extension of the SM gauge group to $SU(3)_C \otimes SU(3)_L \otimes U(1)_X$. We denote this scenario as 3-3-1 model\cite{Pisano:1992bxx,Frampton:1992wt}, which has been extensively studied in the literature since it can provide a mechanism for neutrino masses  \cite{Montero:2000rh, Tully:2000kk,Montero:2001ts,Cortez:2005cp,Cogollo:2009yi,Cogollo:2010jw,Cogollo:2008zc,Dias:2012xp,Okada:2015bxa,Vien:2018otl,carcamoHernandez:2018iel,Nguyen:2018rlb,Pires:2018kaj,CarcamoHernandez:2019iwh,CarcamoHernandez:2019vih,CarcamoHernandez:2020pnh}, allows the inclusion of dark matter candidates \cite{Fregolente:2002nx,Hoang:2003vj,deS.Pires:2007gi,Mizukoshi:2010ky,Ruiz-Alvarez:2012nvg,Profumo:2013sca,Dong:2013ioa,Dong:2013wca,Cogollo:2014jia,Dong:2014wsa,Dong:2014esa,Kelso:2014qka,Dong:2014esa,Mambrini:2015sia,Dong:2015rka,deSPires:2016kkg,Alves:2016fqe,RodriguesdaSilva:2014gbi,Carvajal:2017gjj,Dong:2017zxo,Arcadi:2017xbo,Montero:2017yvy,Huong:2019vej,Alvarez-Salazar:2019cxw,VanLoi:2020xcq,Dutra:2021lto,Oliveira:2021gcw}, can address present meson anomalies \cite{Cogollo:2012ek,Cogollo:2013mga,Buras:2014yna,Buras:2015kwd,Queiroz:2016gif,deMelo:2021ers,Buras:2016dxz,Buras:2021rdg}, the strong CP problem \cite{Pal:1994ba,Dias:2003zt,Dias:2003iq,Montero:2011tg,Dias:2018ddy,Dias:2020kbj}, among others.

Since we have extended the gauge group of the SM, the 3-3-1 models predicts, in general, the existence of extra gauge bosons which potentially modifies many observables. This in turn puts stringent lower bounds in their masses, given the non-observation of such particles in the LHC. Currently, lower bounds for the $Z^{\prime}$ mass is around 4 TeV \cite{Alves:2022hcp}. It should be emphasized that the mass of these extra gauge bosons is directly related to the scale in which the 3-3-1 gauge group is broken to the SM one. Thus, if this breaking occurs at a high scale of hundred of TeVs, for instance, there is no hope to see these particles in the present and near-future colliders. 

Another consequence of the non-observation of the extra gauge bosons is that the present discrepancy in the muon magnetic moment measurement \cite{Muong-2:2021ojo} cannot be addressed by one-loop contributions containing such particles, due to their high masses \cite{deJesus:2020ngn}. However, if the scalar sector of the 3-3-1 model could be light enough, this anomaly could potentially be explained in similarity to the 2HDM \cite{Broggio:2014mna,Chun:2016hzs,Wang:2014sda,Abe:2015oca,Crivellin:2015hha,Chun:2015hsa,Han:2015yys,Ilisie:2015tra,Cherchiglia:2017uwv}. One of the aims of this contribution is to close this gap. 

3-3-1 models have a rich scalar sector, featuring usually three triplets. Given the bounds on the new gauge bosons masses, there are consequences for at least some of the scalars predicted. This can be expected given their common dependence on the vev that breaks the group $SU(3)_C \otimes SU(3)_L \otimes U(1)_X$ to $SU(3)_C \otimes SU(2)_L \otimes U(1)_X$. However, given that at least one scalar was discovered at the LHC with mass around the EW scale, one can wonder if any of the other predicted scalars can have masses below the TeV region. This decoupled scenario can actually be realized rendering an effective 3-3-1 model with the same particle content of the 2HDM \cite{Okada:2016whh,Fan:2022dye}. In this contribution we will study this scenario in great detail, applying current constraints coming from collider searches, B-physics, the W-mass measurement, as well as theoretical constraints such as perturbativity of quartic couplings and stability of the scalar potential. Our analysis will complement and extend the ones presented in \cite{Okada:2016whh,Fan:2022dye}. In particular the inclusion of the theoretical constraints will have a major impact on the allowed enhancement of fermion couplings to the extra scalars. We will also discuss the impact of the recent W-mass measurement by the CDF collaboration \cite{CDF:2022hxs} for the allowed parameter region.

This work is organized as follows: in section \ref{sec:review} we review the scalar sector of the 3-3-1 model, and define its effective version, comparing it to the general 2HDM. In section \ref{sec:pheno} we present the phenomenological constraints we considered, while in section \ref{sec:numerical} we perform our numerical analysis. Finally, we conclude in section \ref{sec:conclusion}. Two appendices are included with further technical details.

\section{Scalar sector of 3-3-1 models and the 2HDM}
\label{sec:review}

In this section we intend to provide a brief review of the scalar sector of 3-3-1 models in comparison to the 2HDM. We begin with the scalar potential of both models, presenting their Yukawa sector afterwards. 

\subsection{Scalar potential}

In the version of 3-3-1 model we are considering, only three triplets $\eta$, $\rho$ and $\chi$ are present. They will be responsible for providing masses for the gauge bosons as well as all charged fermions through the Higgs mechanism. As usual, we choose $\chi$ to perform the first breaking of the gauge group, while both $\eta$, $\rho$  are responsible for the other one. Schematically,
\begin{equation*}
SU(3)_C \otimes SU(3)_L \otimes U(1)_X \xRightarrow{v_{\chi}} SU(3)_C \otimes SU(2)_L \otimes U(1)_X \xRightarrow{v_{\eta},v_{\rho}} SU(3)_C \otimes U(1)_\text{em}     
\end{equation*}
Usually, one adopts
\begin{small}
    \begin{equation}
    \eta=\begin{pmatrix}
    \eta^0 \\ \eta^- \\ \eta^{-A}
    \end{pmatrix},
    \quad
    \rho = \begin{pmatrix}
    \rho^+ \\ \rho^0 \\ \rho^{-B}  
    \end{pmatrix}, 
    \quad
    \chi = \begin{pmatrix}
    \chi^A \\ \chi^{B} \\ \chi^0
    \end{pmatrix},
\end{equation}
\end{small}
where the charges under $U(1)_X$ are
\begin{equation}
    X_{\eta}=-\frac{1}{2} -\frac{\beta_{Q}}{2\sqrt{3}},\quad X_{\rho}=\frac{1}{2}-\frac{\beta_{Q}}{2\sqrt{3}}, \quad X_{\chi}=\frac{\beta_{Q}}{\sqrt{3}}.
\end{equation}
As is well-known, to define a 3-3-1 model one must adopt a specific value for $\beta$, which enters in the charge generator $Q=T_3+\beta_{Q} T_8+ X I$. From the triplets, one obtains seven physical scalars: three neutral CP-even ($h_{1}$,$h_{2}$,$h_{3}$); one neutral CP-odd ($A_{0}$); three charged scalars ($h^{\pm}$, $h^{\pm A}$, $h^{\pm B}$). In this work we are going to consider that all couplings in the scalar potential are real (CP conserving potential), allowing to define CP-even/odd scalars unambiguously. Moreover, the triplets acquire vevs denoted by $v_{\eta}$, $v_{\rho}$, $v_{\chi}$ where $v_{\chi}$ (which performs the first symmetry breaking) is assumed to be much higher than the other two. In similarity to 2HDMs, we will define $\tan\beta=v_{\eta}/v_{\rho}$ and notice that $v=\sqrt{v^{2}_{\rho}+v^{2}_{\eta}}=246\;\text{GeV}$.

In terms of the triplets, the scalar potential is given by (we adopt the convention of \cite{Costantini:2020xrn})
\begin{eqnarray}
\label{eq:Vscalar}
V\left(\eta,\rho,\chi\right) & = &  \mu_{1}^{2}\rho^{\dagger}\rho+\mu_{2}^{2}\eta^{\dagger}\eta+\mu_{3}^{2}\chi^{\dagger}\chi+\lambda_{1}\left(\rho^{\dagger}\rho\right)^{2}+\lambda_{2}\left(\eta^{\dagger}\eta\right)^{2}+\lambda_{3}\left(\chi^{\dagger}\chi\right)^{2} \nonumber\\
 &  & +\lambda_{12}\left(\rho^{\dagger}\rho\right)\left(\eta^{\dagger}\eta\right)+\lambda_{13}\left(\chi^{\dagger}\chi\right)\left(\rho^{\dagger}\rho\right)+\lambda_{23}\left(\eta^{\dagger}\eta\right)\left(\chi^{\dagger}\chi\right)\nonumber \\
  &  & +\zeta_{12}\left(\rho^{\dagger}\eta\right)\left(\eta^{\dagger}\rho\right)+\zeta_{13}\left(\rho^{\dagger}\chi\right)\left(\chi^{\dagger}\rho\right)+\zeta_{23}\left(\eta^{\dagger}\chi\right)\left(\chi^{\dagger}\eta\right)\nonumber\\
 &  & -\sqrt{2}f\epsilon_{ijk}\eta_{i}\rho_{j}\chi_{k}+\textrm{h.c.}
\end{eqnarray}

 The above potential is the most general one as long as $\beta_{Q}\neq \pm 1/\sqrt{3}$. The reason is that, for this choice, two of the triplets will have the same charge under the $U(1)_X$ group allowing for quartic terms with an odd number of the same triplet. This feature can be avoided if, for instance, a $\mathbb{Z}_2$ symmetry is introduced. Since the $\chi$ triplet will be responsible for the breaking of $SU(3)_C \otimes SU(3)_L \otimes U(1)_X$ to $SU(3)_C \otimes SU(2)_L \otimes U(1)_X$, one usually imposes that only this field is odd under the $\mathbb{Z}_2$ just mentioned\footnote{The $\mathbb{Z}_2$ symmetry will be softly broken by the term $\sqrt{2}f\epsilon_{ijk}\eta_{i}\rho_{j}\chi_{k}$. If this term is discarded, one generates a massless CP-odd scalar, see eq. (\ref{eq:massA}).}. In this contribution we will, as far as possible, keep the discussion in general terms which implies that $\beta_{Q}$ will be kept arbitrary. In order to our results to be also applied to the specific choices $\beta_{Q}= \pm 1/\sqrt{3}$, we will implicitly adopt the $\mathbb{Z}_2$ symmetry in this case.

Once the scalar potential is defined, one can study the conditions that the parameters must fulfill in order to respect stability (boundedness from below), perturbativity, and unitarity bounds. In \cite{Costantini:2020xrn} a comprehensive study of these conditions were performed. In particular, they provided expressions which allows one to trade the 13 potential parameters ($\mu_{i}^{2}$, $\lambda_{i}$, $\lambda_{ij}$, $\zeta_{ij}$, $f$) by $\tan\beta$, $v$, $v_{\chi}$, the masses of the seven scalars ($h_{1}$, $h_{2}$, $h_{3}$, $A_{0}$, $h^{\pm}$, $h^{\pm A}$, $h^{\pm B}$) plus three mixing angles that appear when diagonalizing the neutral CP-even mass matrix. Explicitly,
\begin{equation}
    \vec{h} = R^{\text{\tiny{331}}}\vec{H}, 
\end{equation}
where $\vec{H} =(\sqrt{2}\;\Re\;\rho^{0},\sqrt{2}\;\Re\;\eta^{0},\sqrt{2}\;\Re\;\chi^{0})$, $\vec{h}=(h_{1},h_{2},h_{3})$, and 
\begin{equation}
   R^{\text{\tiny{331}}} =
   \begin{pmatrix}
   c_{2}c_{3} & c_{3}s_{1}s_{2}-c_{1}s_{3} & c_{1}c_{3}s_{2}+s_{1}s_{3}\\ 
   c_{2}s_{3} & s_{3}s_{1}s_{2}+c_{1}c_{3} & c_{1}s_{3}s_{2}-s_{1}c_{3}\\ 
   -s_{2} & c_{2}s_{1} & c_{1}c_{2}
   \end{pmatrix}.
\end{equation}
We are adopting the usual convention $c_{i}=\cos(\alpha_{i})$, $s_{i}=\sin(\alpha_{i})$. It can be shown (see for instance \cite{Okada:2016whh,Fan:2022dye}) that the limit  $v_{\chi}>>v_{\rho},v_{\eta}$ implies $\alpha_{2}\sim\alpha_{1}\sim 0$. Thus, only $h_{1}$ and $h_{2}$ can mix, in similarity to the 2HDM. Moreover, $h_{3}$, $h^{\pm A}$, and $h^{\pm B}$ will decouple, rendering an effective scalar potential in $SU(2)_L\otimes U(1)_X$
\begin{eqnarray}
V_{eff}(\Phi_\eta,\Phi_\rho)&=&\mu^2_\rho\Phi^\dagger_\rho\Phi_\rho+ \mu^2_\eta \Phi_\eta^\dagger\Phi_\eta +\lambda_1 (\Phi_\rho^\dagger\Phi_\rho)^2+\lambda_2 (\Phi_\eta^\dagger\Phi_\eta)^2\nonumber \\&+&\lambda_{12}(\Phi_\rho^\dagger\Phi_\rho)(\Phi_\eta^\dagger\Phi_\eta)+
\zeta_{12}(\Phi^T_\eta\varepsilon\Phi_\rho)^\dagger(\Phi_\eta^T\varepsilon\Phi_\rho)
\nonumber\\&+&
\{\mu_{\eta\rho}^{2}\Phi^T_\eta\varepsilon \Phi_\rho +
\text{h.c.}\}\,,
\label{n2}
\end{eqnarray}
where $\varepsilon=i\tau_{2}$ to comply with scalar products in the $SU(2)$ group and the doublets have hypercharge $Y=(-1/2,1/2)$ respectively as below
\begin{equation}
\Phi_\eta=\left(\begin{array}{cc}
\eta^0\\ -\eta^-
\end{array}
\right),\quad \Phi_\rho=\left(\begin{array}{cc}
\rho^+\\ \rho^0
\end{array}\right).
\end{equation}

It is fruitful to consider the general 2HDM at this point. In this model, one introduces two doublets ($\Phi_{1}$, $\Phi_{2}$) under $SU(2)_L$, with the same hypercharge under $U(1)_{Y}$ which acquire the vevs $v_{1}$, and $v_{2}$ respectively. Usually, it is defined $\tan\beta=v_{2}/v_{1}$ and we notice that $v=\sqrt{v^{2}_{1}+v^{2}_{2}}=246\;\text{GeV}$. Explicitly, the doublets can be expressed as
\begin{small}
    \begin{equation}
    \Phi_{i}=\begin{pmatrix}
    \Phi_{i}^+ \\ \Phi_{i}^0   
    \end{pmatrix}.
\end{equation}
\end{small}

In terms of the doublets, the most general scalar potential is\;\cite{Gunion:2002zf}
\begin{align}
\label{eq:Vscalar2HDM}
V\left(\Phi_{1},\Phi_{2}\right)  = &   m_{11}^{2}\Phi_{1}^{\dagger}\Phi_{1}+m_{22}^{2}\Phi_{2}^{\dagger}\Phi_{2}+\frac{\Lambda_{1}}{2}\left(\Phi_{1}^{\dagger}\Phi_{1}\right)^{2}+\frac{\Lambda_{2}}{2}\left(\Phi_{2}^{\dagger}\Phi_{2}\right)^{2} \nonumber\\
 & +\Lambda_{3}\left(\Phi_{1}^{\dagger}\Phi_{1}\right)\left(\Phi_{2}^{\dagger}\Phi_{2}\right)+\Lambda_{4}\left(\Phi_{1}^{\dagger}\Phi_{2}\right)\left(\Phi_{2}^{\dagger}\Phi_{1}\right) + \big[-m_{12}^{2}\Phi_{1}^{\dagger}\Phi_{2} \nonumber\\ &+ \frac{\Lambda_{5}}{2}\left(\Phi_{1}^{\dagger}\Phi_{2}\right)^{2} + \Lambda_{6}\left(\Phi_{1}^{\dagger}\Phi_{1}\right)\left(\Phi_{1}^{\dagger}\Phi_{2}\right)+\Lambda_{7}\left(\Phi_{2}^{\dagger}\Phi_{2}\right)\left(\Phi_{1}^{\dagger}\Phi_{2}\right)+\textrm{h.c.}\big]
\end{align}

A $\mathbb{Z}_2$ symmetry can be considered to render $\Lambda_{6}=\Lambda_{7}=0$, while the term $m_{12}^{2}$ is a soft-breaking term. This model features four physical scalars: two neutral CP-even ($h$, $H$); one neutral CP-odd ($A$); one charged scalar ($H^{\pm}$). In \cite{Gunion:2002zf} it is shown how to trade seven of the parameters in the potential ($m_{ii}^{2}$, $m_{12}^{2}$, $\lambda_{i}$ where $i=1\cdots4$) by $\tan\beta$, $v$, the masses of the four physical scalars and the mixing angle that appears when diagonalizing the neutral CP-even mass matrix. Explicitly, 
\begin{equation}
    \vec{h} = R^{\text{\tiny{2HDM}}}\vec{H},
\end{equation}
where $\vec{H} =(\sqrt{2}\text{Re}\Phi_{1}^{0},\sqrt{2}\text{Re}\Phi_{2}^{0})$, $\vec{h}=(H,h)$, and 
\begin{equation}
   R^{\text{\tiny{2HDM}}} =
   \begin{pmatrix}
   c_{\alpha} & -s_{\alpha}\\ 
   s_{\alpha} & c_{\alpha}
   \end{pmatrix}.
\end{equation}

We emphasize that it is not possible to trade all potential parameters from $V\left(\Phi_{1},\Phi_{2}\right)$ by physical parameters as we did for $V\left(\eta,\rho,\chi\right)$. Namely, we retain $\Lambda_{5}$, $\Lambda_{6}$, and $\Lambda_{7}$. This feature is expected since, by comparing $V\left(\Phi_{1},\Phi_{2}\right)$ with $V_{eff}(\Phi_\eta,\Phi_\rho)$, one notices that in the latter only four quartic couplings are present. It is a consequence of the 3-3-1 symmetry, that will play an important role in the phenomenology. To be concrete, even if the usual $\mathbb{Z}_2$ symmetry is assumed in the scalar potential (which renders $\Lambda_6=\Lambda_7=0$), in the 2HDM $\Lambda_{5}$ is a free parameter. This implies, for instance, that the coupling $\lambda_{hAA}$ can always be set to zero by a particular choice of $\Lambda_5$. The same does not occur in the 3-3-1 model. Explicitly, in the alignment limit ($\beta-\alpha=\pi/2$) one obtains for the 2HDM
\begin{equation}
    \lambda_{hAA}= i\left[ 2\Lambda_{5} v - \frac{m_{h}^{2}}{v}\right]\,,
\end{equation}
while the result of the 3-3-1 model is obtained by setting $\Lambda_{5}=0$ (we are ignoring terms $\mathcal{O}(v_{\chi}^{-1})$)\footnote{In our numerical analysis we consider the full expression, showing that the approximation above is excellent}. Another interesting triple scalar couplings is $\lambda_{hH^{+}H^{-}}$ which, in the alignment limit, is given by
\begin{align}
\label{eq:lhHpHm}
    \lambda_{hH^{+}H^{-}}=
    \begin{cases}
    i\left( 2\frac{m_{A_0}^{2}-m_{h^{\pm}}^{2}}{v}  - \frac{m_{h}^{2}}{v}+ 2\Lambda_{5} v\right) ,\quad&\mbox{2HDM}\\
    i\left( 2\frac{m_{A_0}^{2}-m_{h^{\pm}}^{2}}{v} - \frac{m_{h}^{2}}{v}\right) .\quad&\mbox{331}
    \end{cases}
\end{align}
Once again, this coupling can be suppressed/enhanced in the 2HDM by a suitable choice of $\Lambda_{5}$, while for the 3-3-1 model the coupling is controlled only by the splitting among the masses of the charged and CP-odd scalars.

\subsection{Yukawa sector}

We consider now the Yukawa sector of the 3-3-1 model in comparison to the 2HDM. As it is well-known, we have some choices for the leptonic fields, which will define the quark sector by anomaly cancellation requirements. In a generic way we define
\begin{small}
    \begin{equation}
    L_{i}=\begin{pmatrix}
     e_{i} \\ -\nu_{i} \\ E_{i} 
    \end{pmatrix}_{L},
\end{equation}
\end{small}
\noindent
where $E_{iL}$ can be either a new field or the charge conjugate of the charged lepton field. As can be seen, we define the leptonic fields to be antitriplets under 3-3-1. The right-handed fields are chosen as singlets. The electric charge of the $E_{i}$ fields (both left and right) is given by
\begin{equation}
    Q_{E_{i}} = -\frac{1}{2}+\frac{\sqrt{3}\beta_{Q}}{2}\,.
\end{equation}

Once the leptonic fields are defined, we must choose how to distribute the quarks fields. Given the anomaly requirement, one of the quark families must also be an antitriplet while the other two will be triplets. As usual, we choose the third family, amounting to
\begin{small}
    \begin{equation}
    q_{1}=\begin{pmatrix}
     u \\ d \\ D 
    \end{pmatrix}_{L}\,,
    \quad\quad
    q_{2}=\begin{pmatrix}
     c \\ s \\ S 
    \end{pmatrix}_{L}\,,
     \quad\quad
    q_{3}=\begin{pmatrix}
     b \\ -t \\ T 
    \end{pmatrix}_{L}\,.
\end{equation}
\end{small}
Once again, the right-handed fields are all singlets. The electric charges of the $D,S,T$ fields (both left and right-handed) are given by
\begin{equation}
    Q_{D,S} = \frac{1}{6}-\frac{\sqrt{3}\beta_{Q}}{2}\,,\quad\quad
    Q_{T} = \frac{1}{6}+\frac{\sqrt{3}\beta_{Q}}{2}\,.
\end{equation}
Given the particle content for the fermionic fields, we can write the Yukawa lagrangian as below\footnote{For the choice $\beta_{Q}= \pm 1/\sqrt{3}$, we will consider the fields $D_{R}$, $S_{R}$, $T_{R}$, and $E_{R}$ odd under the $\mathbb{Z}_2$ introduced for this case.}
\begin{align}
-\mathcal{L}_{\rm{Yuk}}^{q} &= y_{ij}^{u}\bar{q}_{iL}\eta u_{jR} + y_{3j}^{u}\bar{q}_{3L}\rho^{*} u_{jR} + y_{ij}^{d}\bar{q}_{iL}\rho d_{jR} + y_{3j}^{d}\bar{q}_{3L}\eta^{*} d_{jR} \nonumber\\&+ y_{i}^{D}\bar{q}_{iL}\chi D_{R} + y_{i}^{S}\bar{q}_{iL}\chi S_{R} +y^{T}\bar{q}_{3L}\chi^{*} T_{R} + \text{h.c.}\,,\\
-\mathcal{L}_{\rm{Yuk}}^{l} &= y_{mn}^{e}\bar{L}_{mL}\eta^{*} e_{nR} + y_{mn}^{E}\bar{L}_{mL}\chi^{*} E_{nR}+ \text{h.c.}\,,
\end{align}
where $i=\{1,2\}$, while the other indexes go from 1 to 3.

For a clearer comparison with the 2HDM, we will rewrite the lagrangian above as
\begin{align}\label{eq:lag}
-\mathcal{L}_{\rm{Yuk}}^{q} &= y_{1j}^{u}\left(\bar{u}_{L}\eta^{0}+\bar{d}_{L}\eta^{-}\right) u_{jR} + y_{2j}^{u}\left(\bar{c}_{L}\eta^{0}+\bar{s}_{L}\eta^{-}\right) u_{jR} + y_{3j}^{u}\left(\bar{b}_{L}\rho^{-}-\bar{t}_{L}\rho^{0}\right) u_{jR}\nonumber\\ 
&+ y_{1j}^{d}\left(\bar{u}_{L}\rho^{+}+\bar{d}_{L}\rho^{0}\right) d_{jR} + + y_{2j}^{d}\left(\bar{c}_{L}\rho^{+}+\bar{s}_{L}\rho^{0}\right) d_{jR} + y_{3j}^{d}\left(\bar{b}_{L}\eta^{0}-\bar{t}_{L}\eta^{+}\right) d_{jR}  \nonumber\\&+ \mbox{terms with $D$, $S$, $T$}+ \text{h.c.}\,,\\
-\mathcal{L}_{\rm{Yuk}}^{l} &= y_{mn}^{e}\left(\bar{e}_{mL}\eta^{0}-\bar{\nu}_{mL}\eta^{+}\right)e_{nR}  +\mbox{terms with $E_{i}$}+ \text{h.c.}
\end{align}
In the format above, it is obvious that $\eta$ will generate the masses  for the charged leptons, the third generation of down-type quarks as well as for first and second generation of up-quarks, while $\rho$ will provide masses for all the remaining charged fermions that are present in the SM. We will work in the limit in which the new fermions do not mix with the SM ones.

We consider that the initial Yukawa lagrangian was written in a basis in which the charged leptons are diagonal. Thus
\begin{equation}
    y_{ii}^{e}=\frac{\sqrt{2}m^{e}_{i}}{v}\frac{v}{v_{\eta}}=\frac{\sqrt{2}m^{e}_{i}}{v}\frac{1}{s_{\beta}},
\end{equation}
where $m^{e}_{i}=\{m_{e},m_{\mu},m_{\tau}\}$. For the up and down quarks, we define the matrices $V$ and $U$ that perform their bi-diagonalization respectively as below
\begin{align}\label{eq:yukawau}
   M_{u}\equiv \frac{v_{\eta}}{\sqrt{2}}V^{\dagger}_{u}\begin{pmatrix}
    y_{11}^{u} & y_{12}^{u} & y_{13}^{u} \\
    y_{21}^{u} & y_{22}^{u} & y_{23}^{u} \\
    \frac{-y_{31}^{u}}{\tan\beta} & \frac{-y_{32}^{u}}{\tan\beta} & \frac{-y_{33}^{u}}{\tan\beta} 
    \end{pmatrix}U_{u}&=\begin{pmatrix}
    m_{u} & 0 & 0 \\
    0& m_{c}  & 0 \\
    0 & 0 & m_{t} 
    \end{pmatrix}\,,
\\
   M_{d}\equiv \frac{v_{\eta}}{\sqrt{2}}
    V^{\dagger}_{d}\begin{pmatrix}
    \frac{y_{11}^{d}}{\tan\beta} & \frac{y_{12}^{d}}{\tan\beta} & \frac{y_{13}^{d}}{\tan\beta} \\
    \frac{y_{21}^{d}}{\tan\beta} & \frac{y_{22}^{d}}{\tan\beta} & \frac{y_{23}^{d}}{\tan\beta} \\
    y_{31}^{d} & y_{32}^{d} & y_{33}^{d} 
    \end{pmatrix}U_{d}&=
    \begin{pmatrix}
    m_{d} & 0 & 0 \\
    0& m_{s}  & 0 \\
    0 & 0 & m_{b} 
    \end{pmatrix}\,.\label{eq:yukawad}
 \end{align}
 
As discussed in \cite{Fan:2022dye}, the matrix $V_{u}$ can be chosen freely while the matrix $V_{d}$ will be constructed by the knowledge of the CKM matrix, $V_{\text{\tiny{CKM}}}=V_{u}^{\dagger}V_{d}$. Under the assumption that the only source of CP-violation comes from the CKM matrix, we can parametrize the matrix $V_{u}$ as
\begin{equation}
   V_{u}^{\dagger} =
   \begin{pmatrix}
   1 & 0 & 0\\
   0 & c_{\psi} & s_{\psi}\\
   0 & -s_{\psi} & c_{\psi}
   \end{pmatrix}
   \begin{pmatrix}
   c_{\theta} & 0 & s_{\theta}\\
   0 & 1 & 0\\
   -s_{\theta} & 0 & c_{\theta}
   \end{pmatrix}
   \begin{pmatrix}
   c_{\phi} & s_{\phi} & 0\\
   -s_{\phi} & c_{\phi} & 0\\
   0 & 0 & 1
   \end{pmatrix}.\label{eq:vu}
\end{equation}

It can be noticed that, if the angles $\psi$ and $\theta$ are small, the matrix will be approximately block diagonal. In this scenario, we have   
\begin{equation}
    \sqrt{(y_{31}^{u})^2+(y_{32}^{u})^2+(y_{33}^{u})^2}=\frac{\sqrt{2}m_{t}}{v}\frac{v}{v_{\eta}}\tan\beta=\frac{\sqrt{2}m_{t}}{v}\frac{1}{c_{\beta}}\,.
    \label{eq:top}
\end{equation}

Since $\frac{\sqrt{2}m_{t}}{v}\approx 1$, it is natural to consider the regime $\tan\beta<1$ to fulfill perturbativity of the Yukawa couplings. As it is clear from eqs.(\ref{eq:yukawau}-\ref{eq:yukawad}), none of the usual 2HDM types (types I, II, X, Y) \cite{Branco:2011iw} can be mapped in the 3-3-1 model. Regarding the effective 3-3-1 model considered in this work, one obtains
\begin{align}\label{eq:n2HDM}
-\mathcal{L}_{\rm{Yuk}}^{q} &= y_{mi}^{u}\bar{q}_{mL}\phi_{\eta} u_{iR} + y_{3i}^{u}\bar{q}_{3L}\phi_{\rho}^{*} u_{iR} + y_{mi}^{d}\bar{q}_{mL}\phi_{\rho} d_{iR} + y_{3i}^{d}\bar{q}_{3L}\phi_{\eta}^{*} d_{iR} +  \text{h.c.}\,,\\
-\mathcal{L}_{\rm{Yuk}}^{l} &= y_{mn}^{e}\bar{L}_{mL}\phi_{\eta}^{*} e_{nR} + \text{h.c.}\,,\label{eq:n2HDM2}
\end{align}
where in the equation above only the first two components of $q_{i}$ and $L_{i}$ are considered. In the mass eigenstate basis and in the alignment limit, the Yukawa lagrangian is given by\cite{Fan:2022dye}
\begin{align}
    -\mathcal{L}_{\rm{Yuk}}^{l}&=\frac{1}{v}\bar{l}_{i} m^{e}_{i} l_{i}\left(h + \frac{1}{\tan\beta} H + i \frac{1}{\tan\beta} \gamma_5 A_{0} \right) \nonumber\\&+ \left(\frac{\sqrt{2}}{v}\bar{l}_{i} m^{e}_{i} P_L\nu_{i}\frac{1}{\tan\beta}H^- + \text{h.c.}\right)\,,
\nonumber\\
    -\mathcal{L}_{\rm{Yuk}}^{q}&=\frac{1}{v}\sum_{\varphi=h,H,A_{0}}\sum_{q=u,d}p_\varphi^q\bar{q}\Gamma_\varphi^q  M_q^{\rm diag} P_R q\varphi+ \text{h.c.} \nonumber\\
    &+\frac{\sqrt{2}}{v}\left[ \bar{d}\left( M_L P_L + M_R P_R \right) u\right]H^- + \text{h.c.}\,,
\end{align}
where $ p_h^q = p_H^q = 1,~p_A^q = 2iI_q$ with $I_q = 1/2 (-1/2)$ for $q=u(d)$, $P_{L,R}$ are the projection operators and 
\begin{align}
M_L = -M_d \Gamma_A^d V_{\text{\tiny{CKM}}}^\dagger, \quad M_R = V_{\text{\tiny{CKM}}}^\dagger \Gamma_A^u M_u. \label{eq:mlmr}
\end{align}
The interaction matrices are given by 
\begin{align}
\Gamma_A^u &= -\frac{1}{\tan\beta} + V_u^\dagger \begin{pmatrix}
    0 & 0 & 0 \\
    0& 0  & 0 \\
    0 & 0 & \frac{1}{\tan\beta}+\tan\beta
    \end{pmatrix}V_u , \nonumber\\
\Gamma_A^d &= \tan\beta - V_{\text{\tiny{CKM}}}^\dagger V_{u}^{\dagger} \begin{pmatrix}
    0 & 0 & 0 \\
    0& 0  & 0 \\
    0 & 0 & \frac{1}{\tan\beta}+\tan\beta
    \end{pmatrix}V_u V_{\text{\tiny{CKM}}} ,  \\
\Gamma_H^q &= -\Gamma_A^q ,\quad \Gamma_h^d = 1
\end{align}
In the format above, it is clear that the angle $\phi$ of $V_{u}$ will not appear, see eq. (\ref{eq:vu}). It is instructive to write the element (33) of the matrix $\Gamma_A^u$ explicitly, since it will be responsible for the coupling among the neutral scalars $H$, $A_{0}$ and the top quark
\begin{equation}
    (\Gamma_A^u)_{33}= c_{\psi}^{2} c_{\theta}^{2} \tan\beta -\frac{c_{\psi}^{2} s_{\theta}^{2}+s_{\psi}^{2} }{\tan\beta}\,.
    \label{eq:gau}
\end{equation}
As can be seen, if the mixing angles are small, the coupling will be proportional to $\tan\beta$, being suppressed if $\tan\beta$ is small as we already discussed after eq. (\ref{eq:top}). On the other hand, for general choices of the mixing angles, the coupling will contain a mixture of terms proportional to $\tan\beta$ and $\cot\beta$. Since the Yukawa coupling related to the top quark is already of order one, one expect in this case that $\tan\beta\sim1$.

\section{Phenomenological and theoretical constraints}
\label{sec:pheno}

In this section we describe the constraints we are going to impose in our model, aiming to illustrate the main differences among the 3-3-1 model in the decoupled regime and the 2HDM.

\subsection{Theoretical constraints}

Since we have a model with multiple scalars, one should consider some theoretical constraints such as the stability of the scalar potential, perturbativity of its couplings as well as perturbative unitarity of the scattering matrix. For 2HDM, these conditions are well-known\cite{Maniatis:2006fs,Ginzburg:2005dt}. For the 3-3-1 model, a detailed study with a user-friendly implementation in Mathematica was performed in \cite{Costantini:2020xrn}. In \cite{Sanchez-Vega:2018qje} the stability conditions for the scalar potential were also studied. Since they will play a major role in constraining the scalar masses, we reproduce them below assuming that $\lambda_3\sim\lambda_{i3}\sim\zeta_{i3}\sim0$ ($i=1,2$) 
\begin{equation}\label{sta}
    \lambda_{1}>0,\quad\lambda_{2}>0,\quad\lambda_{12}+2\sqrt{\lambda_1\lambda_2}>0,\quad\lambda_{12}+\zeta_{12}+2\sqrt{\lambda_1\lambda_2}>0.
\end{equation}
We will show in the numerical section that, in the parameter space considered in this work, the assumption $\lambda_3\sim\lambda_{i3}\sim\zeta_{i3}\sim0$ indeed holds. Moreover, it can be easily seen that the stability conditions of eq. (\ref{sta}) are exactly the same for the 2HDM under the mapping $\lambda_{12}\rightarrow\Lambda_3$, $\zeta_{12}\rightarrow\Lambda_4$, and $\Lambda_5=0$.

Finally, when performing the numerical scan for the 3-3-1 model we will use the full result of \cite{Costantini:2020xrn}. For the 2HDM we will use an in-house routine which was checked against the 2HDMC code \cite{Eriksson:2009ws}.

\subsection{$B\rightarrow X_{s} \gamma$}

\begin{figure}[ht!]
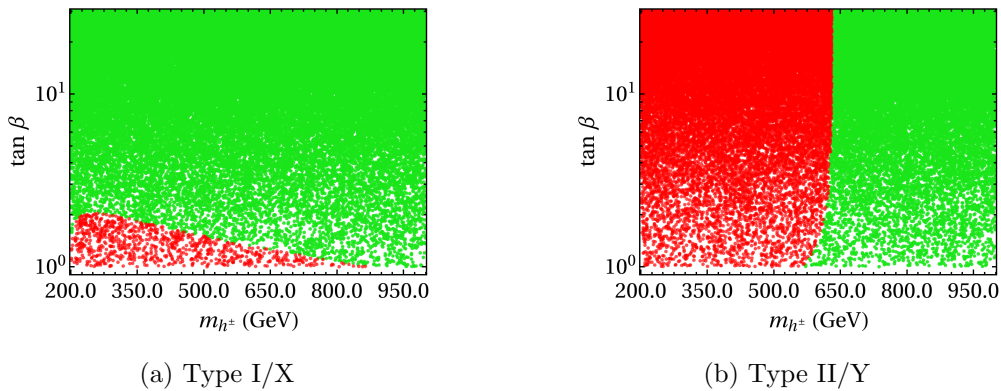

\centering
\begin{subfigure}{.5\textwidth}
  \centering
  \includegraphics[scale=0.35]{bsgI.png}
  \captionsetup{justification=centering}
  \caption{Type I/X}
\end{subfigure}%
\begin{subfigure}{.5\textwidth}
  \centering
  \includegraphics[scale=0.35]{bsgII.png}
  \captionsetup{justification=centering}
  \caption{Type II/Y}
\end{subfigure}
\caption{Allowed (green) and excluded (red) regions in the plane $\tan\beta$ and $m_{h^{\pm}}$ for the 2HDM.}
\label{fig:bs}
\end{figure}

The decay $B\rightarrow X_{s} \gamma$ gives stringent constraints for the parameter space of 2HDM, in particular for the lower bound of the charged scalar mass. This is particularly relevant when the Yukawa couplings of the charged scalar to down-type quarks can be enhanced, which happens in Types II and Y models in the regime with large $\tan\beta$. In our model, for small mixing angles in $V_{u}$, a similar pattern is expected to happen since the couplings to the third family quarks behaves oppositely to each other as in the Types II and Y. For comparison, we have performed an analysis of this observable in Types I/X and II/Y using the analytic formulas of \cite{Enomoto:2015wbn}. Our results can be seen in Fig \ref{fig:bs}, where in red we show the points that are excluded at 95$\%$ C.L. while in green we have the allowed points. Our results are compatible with \cite{Arbey:2017gmh}. Since we are considering in the plots $\tan\beta>1$, the influence of other observables (for instance $B_{s}\rightarrow \mu\mu$) is very marginal. The main result is that in Types II/Y there is a lower bound on the mass of the charged scalar around 600 GeV which is fairly independent of $\tan\beta$. 

\begin{figure}[ht!]
\centering
\includegraphics[scale=0.4]{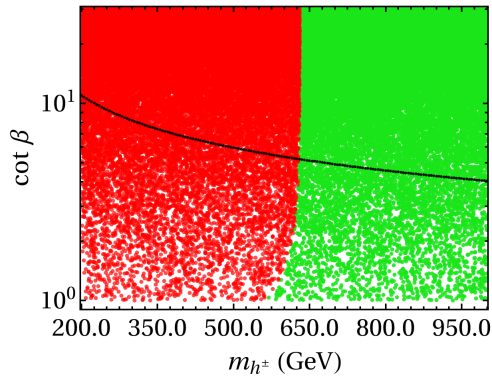}
\caption{\label{fig:bs331}Allowed (green) and excluded (red) regions in the plane $\cot\beta$ and $m_{h^{\pm}}$ for the effective 3-3-1 model. The black line represents an upper limit for $\cot\beta$, see the text for details}
\end{figure}

For the 3-3-1 effective model, we considered the Yukawa sector given in eqs. (\ref{eq:n2HDM}-\ref{eq:n2HDM2}). Notice that the third family up-quarks couples with the doublet $\phi_{1}$ in the notation of eqs. (\ref{eq:n2HDM}-\ref{eq:n2HDM2}), which is opposite to the usual convention in 2HDM. Therefore, in the effective 3-3-1 model with small mixing angles $\psi$ and $\theta$, we obtain in Fig. \ref{fig:bs331} a result similar to type II/Y with $\tan\beta$ exchanged by $\cot\beta$. 

A interesting feature of the effective 3-3-1 model is that there is an upper limit on $\cot\beta$ if $A$ or $H$ are {\textbf{not}} degenerate in mass with the charged scalar $h^{\pm}$. This comes from perturbativity constraints on the quartic couplings, in particular $\lambda_{2}$ which, in the alignment limit, is given by
\begin{equation}\label{eq:l2}
    \lambda_{2}= \frac{m_{h}^{2}}{2v^2} - \frac{(\Delta_{A h^{\pm}}-\Delta_{H h^{\pm}})(\Delta_{A h^{\pm}}+\Delta_{H h^{\pm}}+2m_{h^{\pm}})\cot^2\beta}{2v^2}\,,
\end{equation}
where $\Delta_{ab}=m_{a}-m_{b}$. As can be seen from the equation, unless $\Delta_{A h^{\pm}}=\Delta_{H h^{\pm}}$, there will a upper limit for $\cot\beta$. In fig.\ref{fig:bs331} we depict in black the upper limit for the case $\Delta_{A h^{\pm}}=-100\;\unit{GeV}$, $\Delta_{H h^{\pm}}=-50\;\unit{GeV}$, which is a favoured region given the new result for the W boson mass as we are going to see. 

\begin{figure}[ht!]
\centering
\includegraphics[scale=0.4]{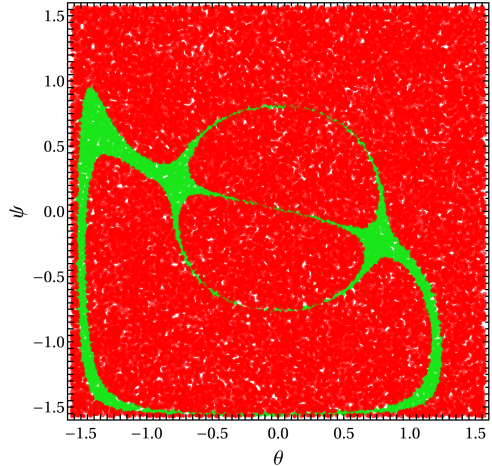}
\caption{\label{fig:bs331Mix} Allowed (green) and excluded (red) regions in the mixing angles plane for $m_{h^{\pm}}=300\;\unit{GeV}$ and $\tan\beta=1$.}
\end{figure}

We conclude this subsection commenting about the case with non-vanishing mixing angles in the matrix $V_u$. As noticed in Ref.\;\cite{Fan:2022dye}, in this scenario one can evade the constraint on $B\rightarrow X_{s} \gamma$ with lower masses for $h^{\pm}$. As a illustrative scenario, we show in fig.\ref{fig:bs331Mix} the allowed regions in the mixing angles plane for $m_{h^{\pm}}=300\unit{GeV}$ and $\tan\beta=1$.

As can be seen, it is necessary to have a strong correlation among the mixing angles to suppress the bound coming from $B\rightarrow X_{s} \gamma$. Finally, as we commented earlier, if $\tan\beta$ is not close to unity, it is possible to have an enhancement on the Yukawa coupling of the neutral scalars to the top quark, which is disfavored by collider constraints (it would increase the production rate of these scalars by gluon fusion, for instance). One could still allow some degree of fine-tuning, since, according to eq. (\ref{eq:gau}), it is possible to suppress this coupling by particular choices of $\psi$, and $\theta$. However, for the scenarios we studied with $\tan\beta=2$, the vast majority of the points that comply with the constraints from $B\rightarrow X_{s} \gamma$ enforce  $(\Gamma_A^u)_{33}>1$. Therefore, in view of this scenario we will restrict ourselves in this work only to the case with null mixing angles.

\subsection{Electroweak constraints}

Regarding electroweak constraints, we consider the S, T, and U parameters. This is particularly relevant given the updated value for the mass of the W-boson presented by the CDF collaboration \cite{CDF:2022hxs}. Using the PDG result for the W mass, one obtains after performing an electroweak fit (hereafter we consider the case $\Delta U=0$) \cite{ParticleDataGroup:2020ssz}
\begin{equation}\label{eq:WPDG}
    \Delta S = 0.00 \pm 0.07, \quad \Delta T = 0.05 \pm 0.06, \quad \text{correlation = } 0.92\,.
\end{equation}

In \cite{deBlas:2022hdk,Lu:2022bgw} a electroweak fit taken into account the CDF result was performed. In \cite{deBlas:2022hdk}, it is considered a combination with previous results for the W mass amounting to
\begin{equation}\label{eq:Wmix}
    \Delta S = 0.086 \pm 0.077, \quad \Delta T = 0.177 \pm 0.070, \quad \text{correlation = } 0.89\,.
\end{equation}

On the other hand, \cite{Lu:2022bgw} considers only the CDF result
\begin{equation}\label{eq:WCDF}
    \Delta S = 0.15 \pm 0.08, \quad \Delta T = 0.27 \pm 0.06, \quad \text{correlation = } 0.93\,.
\end{equation}

In the context of the 2HDM, the correction to the oblique parameters is well-known \cite{Grimus:2008nb}. After the release of the CDF result many works have been performed in the context of 2HDM \cite{Kim:2022xuo,Botella:2022rte,Benbrik:2022dja,Ghorbani:2022vtv,Babu:2022pdn,Song:2022xts} with the main conclusion that the degeneracy in mass of the CP-even and odd scalars with the charged scalar is now disfavoured. We have confirmed these results by using SPheno\cite{Porod:2003um,Porod:2011nf} within the 2HDM, as we are going to show in the numerical section. 

For the 3-3-1 model, we performed two complementary analysis. First we have only the effective version of the 3-3-1 scalar potential given by eq. (\ref{n2}) with the Yukawa sector given by eqs.  (\ref{eq:n2HDM}-\ref{eq:n2HDM2}). In a second step, we used the results presented in \cite{Liu:1993fwa} for the 3-3-1 model. In this reference, it is considered the contributions coming from the heavy spectrum of the 3-3-1 model, in particular the ones related to the charged scalars $h^{\pm A}$, $h^{\pm B}$ as well as the new charged gauge bosons $X, Y$. We checked in our numerical section that these contributions are negligible.

\subsection{Scalar decays}

Since we have an extended scalar sector in our model with respect to the SM, it is natural to consider some extra decays that may be allowed. In particular, for sufficient light scalars, the Higgs boson may itself decay to these new particles. In this case, knowledge of triple scalar couplings is essential and, as we mentioned before, our model in general predicts different patterns than the ones founded in the 2HDM. For instance, for sufficient light CP-odd scalar, the decay $h\rightarrow A_{0}A_{0}$ is open, being actually constant in the alignment regime. Since this decay was not observed, one can immediately exclude a light CP-odd scalar (mass less than $62.5$ GeV) in the region of parameter space of the 3-3-1 model we are considering in this work. Notice that the same does not hold true for the 2HDM, since one has an additional freedom to suppress the coupling by a suitable choice of $\Lambda_5$. A similar reasoning applies for the decay $h \rightarrow HH$. In order to consider similar analysis in a systematic way we make use of HiggsBounds and HiggsSignals \cite{Bechtle:2013wla,Bechtle:2020pkv,Bechtle:2013xfa,Bechtle:2020uwn} when performing our numerical scan.

Finally, we comment on the decay $h\rightarrow \gamma\gamma$ which is only radiately induced, therefore, influenced by all the extra particles in our model. Regarding the scalar sector, the extra contributions come from charged scalars $h^{\pm}$ in the loop\footnote{Regarding the heavier charged scalars and extra gauge bosons, their contributions is negligible in the parameter space we are interested at \cite{Hung:2019jue}.} being controlled by the size of the coupling $\lambda_{h h^{\pm}h^{\pm}}$. As we commented earlier, only if $\Lambda_{5}=0$ the results in our model and 2HDM will be the same. Therefore, we expect some differences on the predictions of this decay that we are going to discuss in the numerical section where we made use of SPheno to obtain the ratio
\begin{equation}
    R_{\gamma\gamma}=\frac{\Gamma(h\rightarrow\gamma\gamma)_{331}}{\Gamma(h\rightarrow\gamma\gamma)_{SM}}\,.
\end{equation}

The ratio above can be analysed in terms of $\kappa$ multipliers, which are provided by the experimental collaborations. Under the assumption that the boson $h$ behaves exactly as the SM Higgs, we obtain the following relation in terms of the $\kappa_\gamma$ multiplier
\begin{equation}
    k_{\gamma}^2=R_{\gamma\gamma}\,.
\end{equation}
Thus, using the value provided by the Atlas collaboration \cite{ATLAS:2019nkf} 
\begin{equation}
    \kappa_{\gamma}=1.00\pm0.06\,,
\end{equation}
we can set the limit
\begin{equation}\label{eq:rgg}
    R_{\gamma\gamma}=1.00\pm0.12
\end{equation}

\subsection{$(g-2)_{\mu}$} 

As described in the introduction, one of the motivations of this work is to study if light scalar particles are allowed in the 3-3-1 model, which may provide a viable explanation to the present $(g-2)_{\mu}$ anomaly ($\amu$) in similarity to the 2HDM. As discussed in \cite{deJesus:2020ngn}, contributions coming from new gauge bosons are suppressed given the present limit on their masses. Using in-house routines based on \cite{Heinemeyer:2004yq,Heinemeyer:2003dq}, we have re-derived their analytic formulas, and confirmed that the contributions are at negligible in the parameter space considered in this work. Regarding the scalar spectrum, we can see in eq. (\ref{eq:lag}) that $h_{3}$ will not couple to the muon line, which prevent their contribution to $\amu$ at one-loop. The remaining one-loop contributions come from the exchange of $h$, $H$, $A_0$, $h^{\pm}$. Since in all cases the coupling to the muon line is proportional to $y_{\mu\mu}^{e}\sim 6\times10^{-4}/s_{\beta}$, only for very low values of $\tan\beta$ can one expect to obtain non-negligible contributions. Nevertheless, two-loop diagrams of Barr-Zee type can surpass this suppression, since in this case the scalar couples to the muon line only once. Moreover, if one has a fermionic loop with a more heavy fermion, the coupling may also be less suppressed. For instance, if one has a loop with the $\tau$ fermion, the coupling will be $y_{\tau\tau}^{e}\sim 10^{-2}/s_{\beta}$. We have performed a general analysis using the results of \cite{Cherchiglia:2016eui} which were recently implemented in the computational tool GM2Calc\cite{Athron:2021evk,Athron:2015rva}. We will discuss our findings in the numerical section. For future reference, we quote the present values for the experimental \cite{Muong-2:2021ojo} and theoretical prediction in the SM \cite{Aoyama:2020ynm} for $\amu$
\begin{equation}
\amu^\text{Exp} = (11659206.1\pm4.1)\times10^{-10}, \quad \amu^\text{SM} = (11659181.0\pm4.3)\times10^{-10}
\end{equation}
which renders a discrepancy by $4.2\sigma$ among the two values. This suggests a BSM contribution as  \begin{equation}
	\amuBSM = (25.1\pm5.9)\times10^{-10}.
    \label{eq:amuBSM}
\end{equation}

\section{Numerical analysis}
\label{sec:numerical}

In this section we perform a focused numerical scan in the parameter space of the 3-3-1 model. This study is complementary to the ones performed in \cite{Hung:2019jue,Okada:2016whh}. At this point is it useful to revise our main assumptions:
\begin{enumerate}
    \item as usual in the literature of 3-3-1 models, we are considering the limit $v_{\chi}>>v_{\rho},v_{\eta}$. This implies that the light spectrum of the 3-3-1 model, in addition to the SM particles, features 3 new scalars ($H$, $A_{0}$, $h^{\pm}$). We identify $h$ as the SM higgs, by considering the alignment limit ($\beta-\alpha=\pi/2$); 
    \item the scalar potential is CP-conserving (all parameters are real). So, there is no extra source of CP violation from the scalar sector;
    \item the exotic fermions do not mix with the SM fermions, and they are decoupled from the theory.
\end{enumerate}

Regarding our scan strategy, we will adopt the physical basis for the scalar potential in the 3-3-1 model, which is characterized by the 13 parameters
\begin{equation*}
    \{\tan\beta, v, v_{\chi}, m_h, m_H, m_{A_{0}}, m_{h^{\pm}}, m_{h_{3}}, m_{h^{\pm A}}, m_{h^{\pm B}}, \alpha_{1}, \alpha_{2}, \alpha_{3}\}
\end{equation*}
Given our assumptions, some of the parameters are chosen as 
\begin{equation*}
    \{v=246 \mbox{ GeV},v_{\chi}=100\mbox{ TeV},  m_{h}=m_{h_{SM}}, \alpha_{1}=\alpha_{2}=0,\alpha_{3}=\alpha=\beta-\pi/2\}
\end{equation*}
leaving us with 7 free parameters whose range we choose as below
\begin{table}[h]
\begin{equation*}
\begin{array}{|c|c|c|c|c|c|c|}
\hline
\tan\beta & m_{H}\, (\unit{GeV})& m_{A_{0}}\,(\unit{GeV})&
m_{h^{\pm}}\,(\unit{GeV})& m_{h_{3}}\,(\unit{TeV}) & m_{h^{\pm A}}\,(\unit{TeV}) & m_{h^{\pm B}}\,(\unit{TeV})
\cr
\hline
10^{-2} \div 1 & 130 \div 1000& 63\div 1000& 600\div 1000&
5\div 10 &
5\div 10 &
5\div 10
\cr
\hline
\end{array}
\end{equation*}
\captionsetup{justification=centering}
\caption{Input parameter ranges.}
\label{table:ranges}
\end{table}

As can be seen, we are adopting $h$ as the lightest CP-even scalar in our model. Recall that the mass of CP-even or odd scalars cannot be lighter than half the mass of $h$, leaving just the window (62.5 GeV - 125 GeV). For $\tan\beta$ we aim to avoid Yukawas larger than unit for the top quark, while the mass for the charged scalar is constraint by the $B\rightarrow X_{s}\gamma$ decay.

For comparison, we also performed a numerical scan in the 2HDM with a soft-broken $\mathbb{Z}_2$ symmetry ($\Lambda_{6}=\Lambda_{7}=0$). We adopted the same ranges for the parameters shared by the 2HDM and 3-3-1 model, implying that the only remaining free parameter is $\Lambda_{5}$ which we vary under the full perturbative range $|\Lambda_{5}|<4\pi$.

We begin by discussing the influence of the theoretical bounds. It will prove fruitful to write the quartic couplings for the 3-3-1 model in the alignment limit
\begin{align}
    \lambda_{1}&= \frac{m_{h}^{2}}{2v^2} + \frac{(m_H^2-m_{A_{0}}^2)\tan^2\beta}{2v^2}\,,\\
    \lambda_{2}&= \frac{m_{h}^{2}}{2v^2} + \frac{(m_H^2-m_{A_{0}}^2)}{2v^2\tan^2\beta}\,,\\
    \lambda_{12}&= \frac{m_{h}^{2}}{2v^2} - \frac{(m_H^2-m_{A_{0}}^2)}{2v^2}\,,\\
    \zeta_{12}&=\frac{2\left(m_{h^{\pm}}^{2}-m_{A_0}^{2}\right)}{v^{2}}\,.
\end{align}

\begin{figure}[ht!]
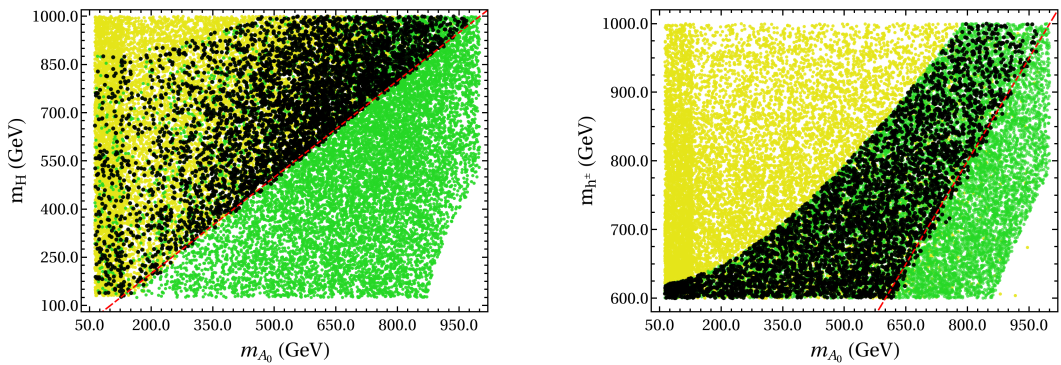

\centering
\begin{subfigure}{.5\textwidth}
  \centering
  \includegraphics[scale=0.4]{mAmH.png}
\end{subfigure}%
\begin{subfigure}{.5\textwidth}
  \centering
  \includegraphics[scale=0.4]{mAmHp.png}
\end{subfigure}
\caption{In both figures, the yellow points comply with the stabiltiy bound, the green points with perturbativity and unitarity, while the black points comply with all theoretical constraints. The dashed red line is for degenerate masses.}
\label{fig:theo}
\end{figure}

Since we require $0<\lambda_{1,2}
<4\pi$, in the regime with $10^{-2}<\tan\beta<1$ the stronger constraint will come from $\lambda_2$. In particular, it will favor the hierarchy $m_{H}>m_{A_0}$ and, for small $\tan\beta$, the two masses will be nearly degenerate. Notice that these conditions for $\lambda_2$ will favor a negative value for $\lambda_{12}$. Finally, since we have the stability constraint,
$\lambda_{12}+\zeta_{12}>-2\sqrt{\lambda_1\lambda_2}$, and $\lambda_{12}$ will be mostly negative, $\zeta_{12}$ will need to be mostly positive which implies that $m_{h^{\pm}}>m_{A_0}$. Therefore, for most of the points in the parameter space we expect the hierarchy $m_{H}>m_{A_0}$ and $m_{h^{\pm}}>m_{A_0}$. We emphasize that the same pattern does not occur in the 2HDM, since we have an additional free parameter, $\Lambda_5$, in that case. This general analysis explains the curves seen in fig. \ref{fig:theo}. Notice that the lower limit on $m_{A_0}$ strongly depends on the value of $m_{h^{\pm}}$.

We briefly comment on the remaining quartic couplings which, in the alignment limit, are given by
\begin{align}
    \lambda_{3}\sim \frac{m_{h_3}^{2}}{2v_{\chi}^2}, \quad    \lambda_{i3}\sim \frac{m_{A_{0}}^{2}}{v_{\chi}^2}, \quad
    \zeta_{13}\sim\frac{2m_{h^{\pm B}}^{2}}{v_{\chi}^2}, \quad
    \zeta_{23}\sim\frac{2m_{h^{\pm A}}^{2}}{v_{\chi}^2}.
\end{align}
As can be seen from the equations above, all the remaining quartic couplings are positive and suppressed given our choice of the parameters. Therefore, they will not play any important role into the theoretical bounds.

\begin{figure}[ht!]
\centering
\includegraphics[scale=0.9]{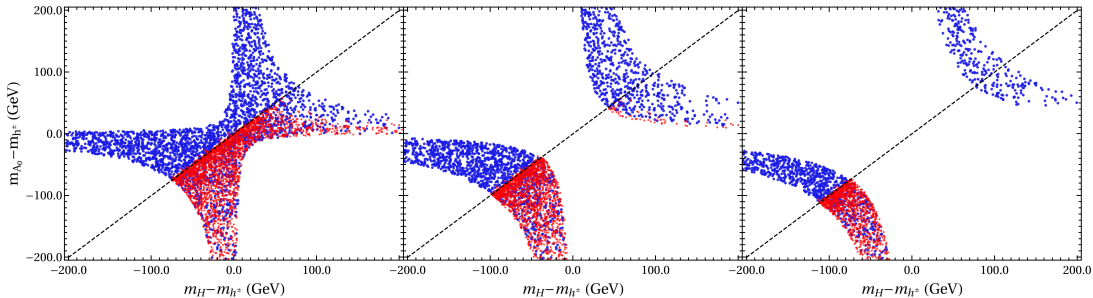}
\caption{\label{fig:st} Difference in mass among the neutral scalars $H$, $A_0$ and the charged scalar $h^{\pm}$ complying at 95$\%$ C.L. with oblique parameters of equations (\ref{eq:WPDG}), (\ref{eq:Wmix}), (\ref{eq:WCDF}) respectively. In blue we show the points in the 2HDM while in red the points of the 3-3-1 model. }
\end{figure}

We consider next the impact of the CDF result for the W mass, which can be seen in fig. \ref{fig:st}, where we show the allowed range at 95$\%$ C.L. for the difference in mass of the neutral scalars $H$, $A_0$ to $h^{\pm}$.

In the left plot we show the case where the W mass according to PDG is used for the prediction of the oblique parameters, eq. (\ref{eq:WPDG}), while in the middle plot a combination with the CDF result is performed, eq. (\ref{eq:Wmix}). For the right plot we consider only the CDF result, eq. (\ref{eq:WCDF}). In all plots we show in blue the points using the 2HDM while the red points correspond to the 3-3-1 model. For the 3-3-1 model we checked that the influence of the heavy spectrum is negligible for the calculation of the S,T parameters. We also checked that, in all cases, the contribution to the U parameter is negligible. As can be seen from the plots, the new result from the CDF collaboration favours a splitting in mass between the charged scalar and the neutral ones. This finding is particularly relevant for the 3-3-1 effective model, since it affects triple scalar couplings such as $\lambda_{h h^{+}h^{-}}$, see eq. (\ref{eq:lhHpHm}). Moreover, contrarily to 2HDM, for most of the points only one of the branches is accessible to the 3-3-1 model due to the favoured hierarchy $m_{h^{\pm}}>m_{A_0}$.

\begin{figure}[ht!]
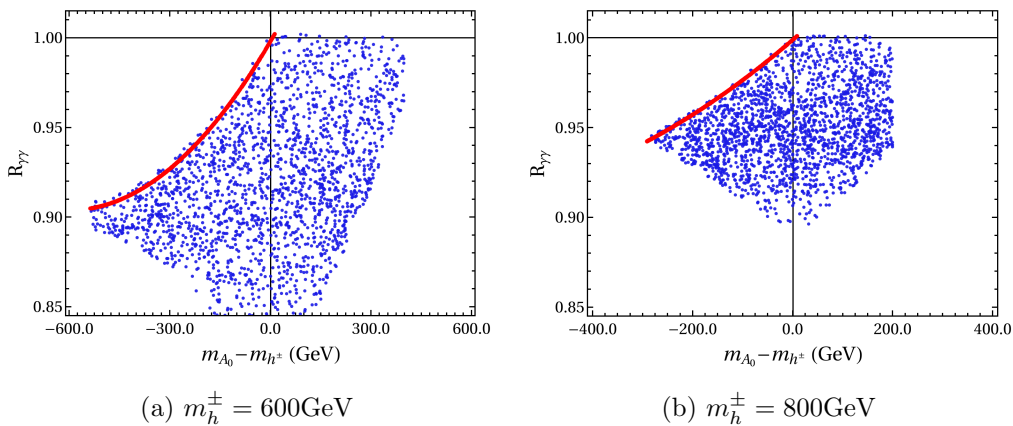

\centering
\begin{subfigure}{.45\textwidth}
  \centering
  \includegraphics[scale=0.4]{hgg600.png}
  \captionsetup{justification=centering}
  \caption{$m_{h}^{\pm}=600\unit{GeV}$}
  \label{fig:fsub2}
\end{subfigure}
\begin{subfigure}{.45\textwidth}
  \centering
  \includegraphics[scale=0.4]{hgg800.png}
  \captionsetup{justification=centering}
  \caption{$m_{h}^{\pm}=800\unit{GeV}$}
  \label{fig:fsub3}
\end{subfigure}
\caption{Ratio $R_{\gamma\gamma}$ as a function of the difference in mass among $A_0$ and the charged scalar $h^{\pm}$. Each plot corresponds to a different choice of $m_{h}^{\pm}$. In all plots, blue points correspond to the 2HDM while red points correspond to the 3-3-1 effective model.
}
\label{fig:hgg}
\end{figure}

In order to illustrate the more predictive behavior of the 3-3-1 effective model, we consider the decay $h\rightarrow \gamma\gamma$ that is directly affected by the coupling $\lambda_{h h^{+}h^{-}}$. Our result can be seen in figure \ref{fig:hgg} where all points comply with theoretical constraints in their respective model. We use the same color code of figure \ref{fig:st}. Both plots show a similar behavior, where the 3-3-1 effective model corresponds to a very precise prediction to $R_{\gamma\gamma}$ for a given choice of $m_{A_0}-m_{h^{\pm}}$. The lower bound for $m_{A_0}$ in the right plot (fig. \ref{fig:fsub3}) is explained by theoretical constraints. Moreover, given the constraint of eq. (\ref{eq:rgg}), the 3-3-1 effective model complies with the experimental value at 1$\sigma$ for all points. 

\begin{figure}[ht!]
\centering
  \includegraphics[scale=0.8]{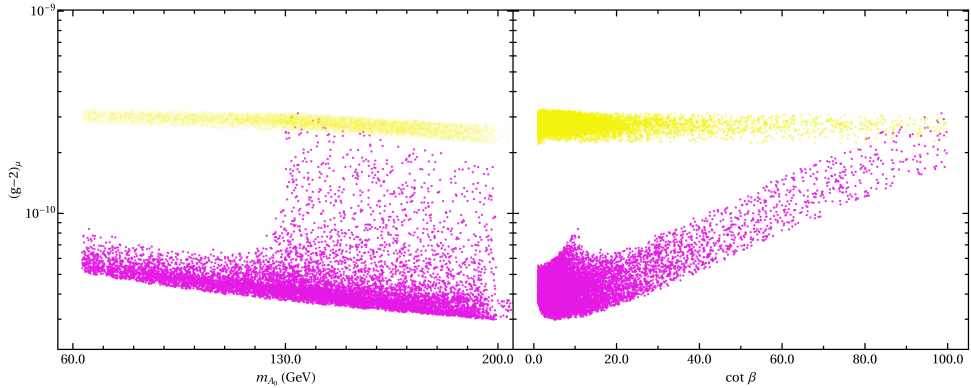}
\caption{Values for $(g-2)_{\mu}$ in the effective 3-3-1 model. In purple we show the fermionic contributions (2-loop Barr-Zee diagrams) plus the 1-loop contributions. In yellow we show the absolute value of the bosonic contribution (which is always negative in the region scanned). All points comply with theoretical constraints.}
\label{fig:g2}
\end{figure}

We consider next the influence of a possible light scalar spectrum in explaining $(g-2)_{\mu}$. In fig. \ref{fig:g2} we show in purple the contribution for the fermionic contribution (2-loop Barr-Zee diagrams) plus the 1 loop contribution (which are negligible for the parameter space considered). In yellow we show the absolute value of the bosonic contribution, which is always negative in the parameter space studied. As can be seen, the value for $(g-2)_\mu$ allowed in the effective 3-3-1 model is too small, see eq. (\ref{eq:amuBSM}). The behavior in the left plot is explained for our choice of $m_{H}>m_{h_{SM}}$. As shown in eq. (\ref{eq:l2}), there is a upper limit on $\cot \beta$ if $m_{H}\neq m_{A_{0}}$. Since to explain $(g-2)_\mu$ an enhancement is required (performed by choosing higher values of $\cot \beta$), one sees that only when $m_{A_{0}}\sim m_{H}$ the fermionic contribution to $(g-2)_\mu$ can increase. However, it is still too small. Moreover, our analysis of the oblique parameters shows that it is not possible to have both $m_{A_{0}}$ and $m_{H}$ far away from $m_{h^{\pm}}\sim 600\unit{GeV}$. Therefore, the region with $m_{H}<130 \unit{GeV}$ and $m_{A_{0}}\sim m_{H}$, which was not included in our scan, is excluded by the oblique parameters. 


\begin{figure}[ht!]
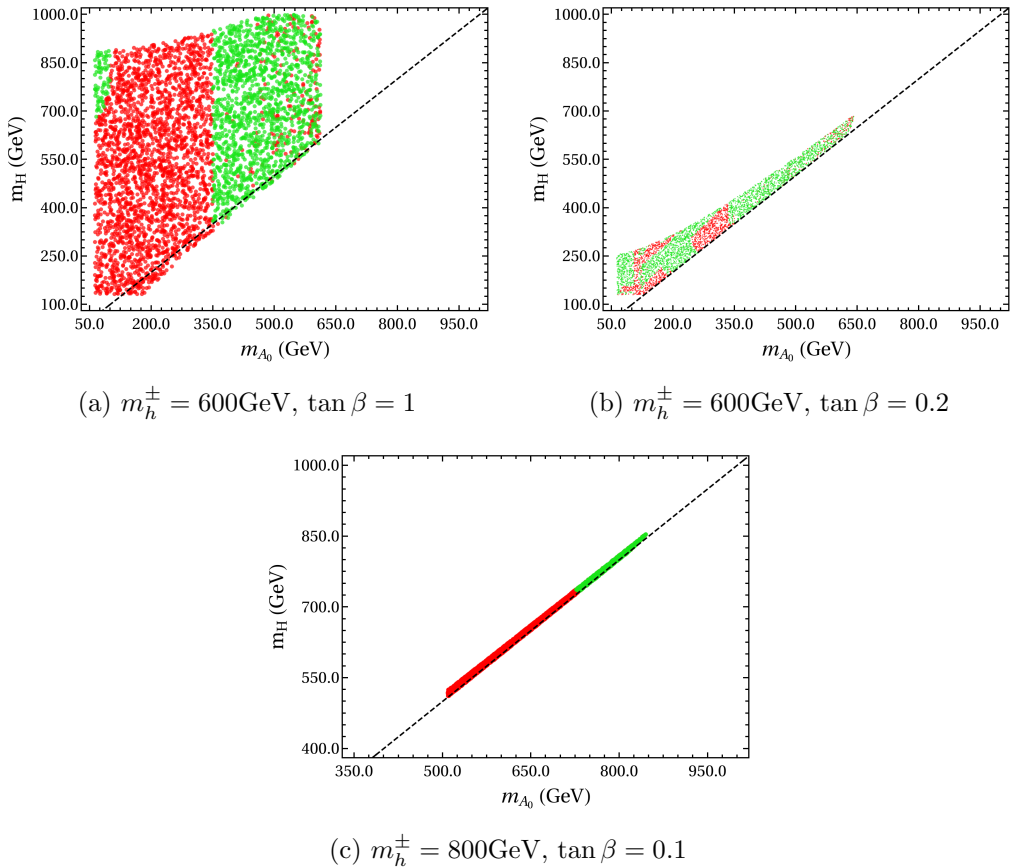

\centering
\begin{subfigure}{.45\textwidth}
  \centering
  \includegraphics[scale=0.4]{mAmH_MHp600TB1.png}
  \captionsetup{justification=centering}
  \caption{$m_{h}^{\pm}=600\unit{GeV}$, $\tan\beta=1$}
  \label{fig:csub1}
\end{subfigure}
\begin{subfigure}{.45\textwidth}
  \centering
  \includegraphics[scale=0.4]{mAmH_MHp600TB5.png}
  \captionsetup{justification=centering}
  \caption{$m_{h}^{\pm}=600\unit{GeV}$, $\tan\beta=0.2$}
  \label{fig:csub2}
\end{subfigure}
\begin{subfigure}{.45\textwidth}
  \centering
  \includegraphics[scale=0.4]{mAmH_MHp800TB10.png}
  \captionsetup{justification=centering}
  \caption{$m_{h}^{\pm}=800\unit{GeV}$, $\tan\beta=0.1$}
  \label{fig:csub3}
\end{subfigure}
\caption{Collider constraints for the allowed range of masses for extra neutral scalars. In red we show the excluded points, while in green the allowed ones at 95$\%$C.L. Values for $m_{h^{\pm}}$ and $\tan\beta$ are chosen as shown in the plots. The black dashed line is for degenerate masses. All points comply with theoretical constraints.}
\label{fig:col600}
\end{figure}

Next, we analyse the constraints coming from searches in colliders. In fig. \ref{fig:col600} we show in green the region that is allowed at 95$\%$ C.L. by using the HiggsBounds\cite{Bechtle:2013wla,Bechtle:2020pkv} and HiggSignals\cite{Bechtle:2013xfa,Bechtle:2020uwn} codes, while in red we show the excluded points. All points comply with theoretical bounds. In order to obtain two-dimensional plots, we have fixed $m_{h^{\pm}}$ and $\tan \beta$ in each plot, allowing $m_{A_{0}}$ and $m_{H}$ to vary in the whole range. As can be seen in fig.\ref{fig:csub1}, for the choice $m_{h^{\pm}}=600 \unit{GeV}$ and $\tan\beta=1$ the region with $m_{A_{0}}<350\unit{GeV}$ is excluded with the exception of a small island with $m_{H}>700\unit{GeV}$ and $m_{A_{0}}<70\unit{GeV}$. For $\tan\beta=0.2$ (fig.\ref{fig:csub2}), there are more regions allowed, and notice that $m_{A_{0}}\sim m_{H}$ as already explained. However, if $\tan\beta$ is decreased further, all points are excluded. It happens already at $\tan\beta=0.1$. This is explained by the hierarchy $m_{A_{0}}\sim m_{H}$ as well as $m_{A_{0}}< m_{h^{\pm}}=600\unit{GeV}$, which implies that $m_{H}< m_{h^{\pm}}=600\unit{GeV}$. Since this range of mass is accessible at the LHC and smaller $\tan\beta$ implies in an enhancement of the Yukawa couplings to the third family down quarks and leptons, the constraints are quite strong. For the case of $m_{h^{\pm}}=800\unit{GeV}$ and $\tan\beta>0.2$ the influence of collider searches is very small, after considering theoretical bounds. Recall that in this case there is a lower bound on $m_{A_{0}}$ of around $500\unit{GeV}$ as can be seen in fig. \ref{fig:theo} or fig.\ref{fig:hgg}. For decreasing values of $\tan\beta$ the constraints become stronger, as can be seen in fig.\ref{fig:csub3} where $\tan\beta=0.1$ were chosen. In this case, masses lower than 720 GeV are excluded. 

\begin{figure}[ht!]
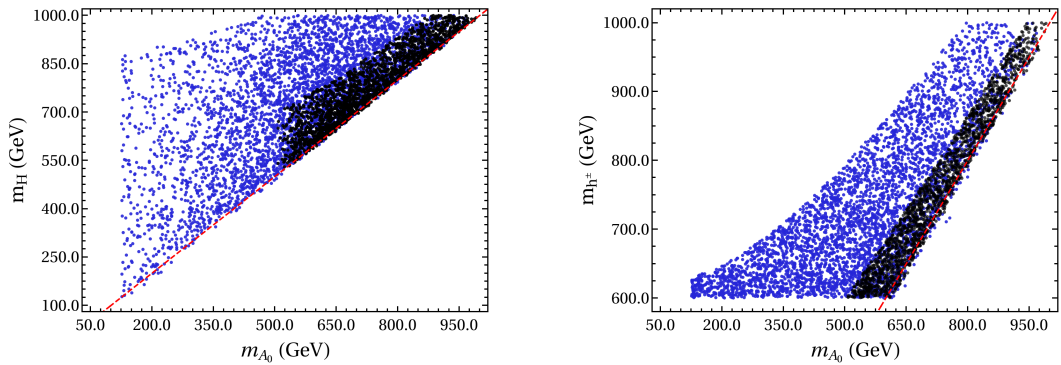

\centering
\begin{subfigure}{.5\textwidth}
  \centering
  \includegraphics[scale=0.4]{mAmHSqrt.png}
  \label{fig:sub1}
\end{subfigure}%
\begin{subfigure}{.5\textwidth}
  \centering
  \includegraphics[scale=0.4]{mAmHpSqrt.png}
  \label{fig:sub2}
\end{subfigure}
\caption{In both figures, blue points complies with theoretical constraints with the choice $\lambda_{\max}=4\pi$, while for the black points we choose $\lambda_{\max}=\sqrt{4\pi}$. The dashed red line is for degenerate masses.}
\label{fig:theo2}
\end{figure}

Next, we show the impact of the maximum value chosen for the pertubative quartic couplings. In fig. \ref{fig:theo2} we show in blue the points that comply with all theoretical constraints using the choice $\lambda_{\max}=4\pi$ while in black we choose $\lambda_{\max}=\sqrt{4\pi}$. An can be seen, in the second case there is a much stronger correlation among $m_{A_{0}}$ and $m_{h^{\pm}}$, implying that $m_{A_{0}}>500\unit{GeV}$. The correlation among $m_{A_{0}}$ and $m_{H}$ is also stronger, for instance, if $m_{A_{0}}=500\unit{GeV}$, $m_{H}$ can be at most of order of 700 GeV. In the previous case, $m_{H}$ reached the ceiling of the chosen range for the scan.

\begin{figure}[ht!]
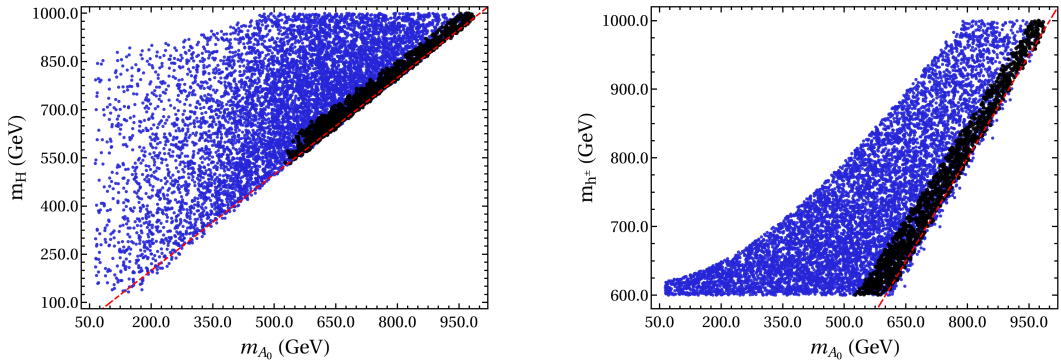

\centering
\begin{subfigure}{.5\textwidth}
  \centering
  \includegraphics[scale=0.4]{mAmHRGE.png}
\end{subfigure}%
\begin{subfigure}{.5\textwidth}
  \centering
  \includegraphics[scale=0.4]{mAmHpREG.png}
\end{subfigure}
\caption{In both figures, blue points complies with theoretical constraints at the EW scale, while for the black points we choose the scale $v_{\chi}=100\;\unit{TeV}$. The dashed red line is for degenerate masses.}
\label{fig:theo3}
\end{figure}

Finally, in light of the later analysis on the  impact of the perturbativity of quartic couplings, we consider the renormalization group equation (RGE) evolution of the quartic couplings from the EW scale up to 100 TeV, which is the scale of $v_{\chi}$. In our analysis, we will consider that the input parameters presented in table~\ref{table:ranges} are defined at the EW scale, which will allows us to obtain the quartic couplings also at this scale. We impose the theoretical bounds already discussed (stability of the scalar potential, perturbativity of the quartic couplings as well as perturbative unitarity of the scattering matrix), which will render the surviving sample depicted as black points in fig.~\ref{fig:theo} where we also adopted $\lambda_{\rm{max}}=4\pi$. We then perform the RGE evolution (further details can be found in the appendix~\ref{ap:rge}) until the scale $v_{\chi}=100\;\unit{TeV}$. We once again apply the theoretical bounds at this scale (in particular we adopt $\lambda_{\rm{max}}=4\pi$), and collect the surviving sample. Similar analysis, in the context of the 2HDM, were performed for instance in \cite{Chakrabarty:2014aya,Chowdhury:2015yja}. Our results can be seen in fig.~\ref{fig:theo3}, where we notice that there is a strong correlation among the masses of the distinct scalars, and we obtain the lower bound $m_{A_{0}}>500\unit{GeV}$.

At this point it is worth to summarize our main findings for a light scalar spectrum for the effective 3-3-1 model considered in this work:
\begin{enumerate}
    \item the constraint from $B\rightarrow X_{s} \gamma$ requires that the mass of the charged scalar is at least 600 GeV. The same statement holds for Type II/Y 2HDM; \item Theoretical constraints such as perturbativity of quartic couplings and stability of the scalar potential will favor the hierarchies $m_{h^{\pm}}>m_{A_0}$ and $m_H>m_{A_{0}}$. For the 2HDM, there is no favored hierarchy;
    \item The CDF result for the W mass disfavors degenerate masses for the scalars. Given the hierarchy in the 3-3-1 effective model, it implies that both $H$ and $A_0$ will be lighter than $h^{\pm}$. For the 2HDM, $H$ and $A_0$ can also be heavier than $h^{\pm}$;
    \item By considering the CDF result for the W mass, one can predict that the diphoton decay of the SM-like scalar normalized to the SM value is always less than unity. For the 2HDM no prediction can be made;
    \item $(g-2)_{\mu}$ cannot be explained in the effective 3-3-1 model, since the region with the lightest scalar does not allow enhancement due to perturbativity of the scalar quartic couplings. Moreover, $m_{A_0}>63\unit{GeV}$. For the 2HDM, not only can $A_0$ be lighter (by a suitable choice of $\Lambda_5$) as there is no constraint on the enhancement;
    \item If the maximum value of the quartic couplings is decreased to $\sqrt{4\pi}$, one obtains a lower bound on $m_{A_{0}}\sim 500\unit{GeV}$. A similar lower bound occurs when we apply RGE evolution of the quartic couplings to our sample, imposing theoretical bounds (in particular $\lambda_{\rm max}=4\pi$) both at the EW scale and at the $v_{\chi}=100\unit{TeV}$ scale. Finally, collider constraints can impose strong constraints for $m_{A_{0}}$, depending on the value of $\tan\beta$. 
\end{enumerate}

\begin{figure}[ht!]
\centering
\begin{subfigure}{.95\textwidth}
  \includegraphics[scale=0.8]{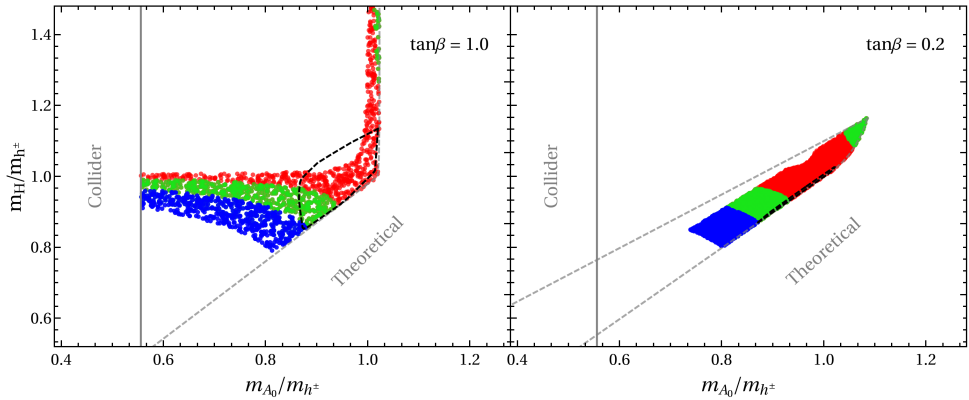}
  \caption{$m_{h}^{\pm}=600\unit{GeV}$}
\end{subfigure}
\begin{subfigure}{0.95\textwidth}
  \includegraphics[scale=0.8]{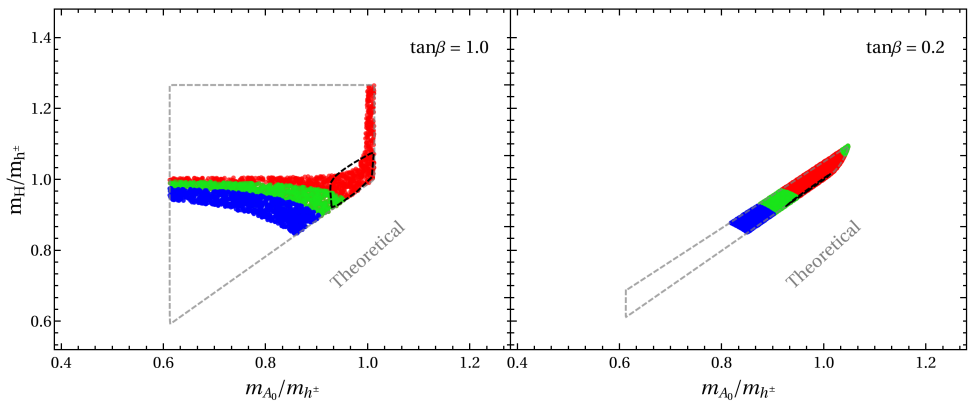}
  \caption{$m_{h}^{\pm}=800\unit{GeV}$}
  \end{subfigure}
  \caption{Summary plots when all constraints are enforced. Outside the dashed lines, the region is excluded by theoretical bounds. The gray dashed lines correspond to imposing $\lambda_{\max}=4\pi$ at the EW scale while the black dashed lines correspond to imposing $\lambda_{\max}=4\pi$ both at the EW scale and at the $v_{\chi}=100\unit{TeV}$ scale. The left region from the continuous gray line is excluded by collider bounds. The mass of the charged scalar and $\tan\beta$ are chosen as indicated in the plots.}
\label{fig:summary}
\end{figure}

We conclude this section with fig.\ref{fig:summary} which illustrates allowed regions of $m_{A_{0}}/m_{h^{\pm}}$ and $m_{H}/m_{h^{\pm}}$ for chosen values of $m_{h^{\pm}}$ and $\tan\beta$, when all constraints are enforced. For each plot, the different regions correspond to the three values of the electroweak oblique parameters. The red points correspond to eq. (\ref{eq:WPDG}), the green points correspond to eq. (\ref{eq:Wmix}), and the blue points correspond to eq. (\ref{eq:WCDF}). Regarding theoretical bounds, we consider two scenarios. In the first (more conservative) we just adopt $\lambda_{\max}=4\pi$, which is shown as dashed gray lines in the plot. The second scenario corresponds to bounds coming from our RGE analysis, which stands for imposing $\lambda_{\max}=4\pi$ both at the EW scale and at the $v_{\chi}=100\unit{TeV}$ scale. This case is represented as black dashed lines. As can be seen in the plot, even for the more conservative case, the mass of the CP-odd neutral scalar cannot be smaller than 55$\%$ of $m_{h^{\pm}}$, while for the CP-even scalar this limit increases to 75$\%$. For the case of $m_{h^{\pm}}=600\unit{GeV}$, this lower bound for $m_{A_{0}}$ is enforced by collider constraints while for $m_{h^{\pm}}=800\unit{GeV}$ the theoretical constraints are responsible for this behavior. For the case of $m_{H}$, the lower bound is due to electroweak oblique parameters. When the RGE analysis is considered, the available parameter space is significantly reduced. In this case, the mass of the CP-odd neutral and CP-even scalar cannot be smaller than 85$\%$ of $m_{h^{\pm}}$. 
 
It is interesting to notice that, regardless of the mass of $h^{\pm}$, if only the CDF result is considered when computing the electroweak oblique corrections (blue points), there is an island for the allowed masses of the neutral scalars which are both lighter than $h^{\pm}$. For smaller values of $\tan\beta$, the region is even more compressed. However, by considering the RGE analysis, we notice that this region is completely excluded for $m_{h^{\pm}}=800\unit{GeV}$ while for $m_{h^{\pm}}=600\unit{GeV}$ there is a very small surviving region. Adopting a more conservative analysis in which the CDF result is combined with previous results (green points), we notice that the RGE analysis allows only the region in which the masses of the neutral scalars are at the same order or lighter than $h^{\pm}$. These findings imply that, if a charged scalar is found, for instance, at the LHC, the effective 3-3-1 model {\textbf{predicts}} the mass region of the other scalars which are necessarily lighter or at most of the same order of $h^{\pm}$. In this scenario, the experimental collaboration could focus on the predicted mass regions and refute (or confirm) the 3-3-1 effective model.

\section{Conclusion}
\label{sec:conclusion}

The 3-3-1 model, proposed in the 90's \cite{Pisano:1992bxx,Frampton:1992wt}, remains an appealing candidate for extending the Standard Model. It predicts the number of fermion generations to be a multiple of three, which shed some light on the current pattern observed in nature. It can also explain the smallness of neutrino masses as well as the strong CP problem. From a more phenomenological point of view, the current collider constraints set the masses of all the new gauge bosons to be at the TeV range or higher, the specific limit depending on the version of the 3-3-1 model considered. More dramatically, since the masses of the extra gauge bosons are directly connected to the vev of the first gauge symmetry breaking, they can be completely non-detectable in present and near-future colliders if the symmetry breaking occurs at a scale of hundred of TeVs, for instance. However, even in this discouraging scenario, the model allows some of the extra scalars to be light, at the sub-TeV range, resembling a two-higgs doublet model (2HDM). 

In this contribution we have studied in detail this effective realization of the 3-3-1 model, aiming to discuss its differences to a general 2HDM. We found that there is a lower and upper limit for the mass of the neutral scalars that strongly depends on the mass of the charged scalar. Thus, if the latter is found at the LHC, for instance, the effective 3-3-1 model predicts the mass range of the other scalars. This is in clear contrast to the general 2HDM, in which a upper limit for the masses cannot be enforced. Moreover, the predicted mass range also depends on the W-mass through the oblique parameters, opening the possibility to support (or not) the CDF result if new scalars are found and their masses determined. Finally, we confirmed that the current deviation for the muon magnetic moment cannot be addressed by 3-3-1 models, by explicitly evaluating its contribution in the effective 3-3-1 model. This analysis complements the one performed in \cite{deJesus:2020ngn}.  

\acknowledgments
We gratefully acknowledge enlightening discussions with Dominik Stöckinger, Hyejung Stöckinger-Kim, and Vicente Pleitez. 
A.C. acknowledges support from National Council for Scientific and Technological Development – CNPq through project 166523\slash 2020-8.  O.L.G.P. were thankful for the support of FAPESP funding Grant 2014/19164-6. and CNPq grant 306565/2019-6. This study was financed in part by the Coordenação de Aperfeiçoamento de Pessoal de Nível Superior - Brasil (CAPES) - Finance Code 001.

\appendix

\section{Inversion relations for the parameters of the scalar potential} 

Using the results provided in \,\cite{Costantini:2020xrn} one can obtain expressions for the parameters of the scalar potential in the effective 3-3-1 model. By performing the choices ($\alpha_{2}=\alpha_{1}=0$; $\alpha_{3}=\alpha$; $m_{h_{2}}=m_{h}$; $m_{h_{1}}=m_{H}$; $m_{A_{0}}=m_{A}$; $m_{h^{\pm}}=m_{H^{\pm}}$), one obtains in the limit $v_{\chi}\gg v_{\rho},v_{\eta}$
\begin{align}
    \lambda_{1}&= \frac{m_{H}^{2}c_{\alpha}^{2}+m_{h}^{2}s_{\alpha}^{2}-m_{A}^{2}s_{\beta}^{2}}{2\;v^{2}c_{\beta}^{2}};\\
    \lambda_{2}&= \frac{m_{H}^{2}s_{\alpha}^{2}+m_{h}^{2}c_{\alpha}^{2}-m_{A}^{2}c_{\beta}^{2}}{2\;v^{2}s_{\beta}^{2}};\\
    \lambda_{12}&=\frac{\left(m_{H}^{2}-m_{h}^{2}\right)s_{\alpha}c_{\alpha} + m_{A}^{2}s_{\beta}c_{\beta}}{v^{2}s_{\beta}c_{\beta}};\\
    \zeta_{12}&=\frac{2\left(m_{H^{\pm}}^{2}-m_{A}^{2}\right)}{v^{2}};\\
    fv_{\chi}&=m_{A}^{2}c_{\beta}s_{\beta};\label{eq:massA}\\
    \mu_{\rho}^{2} &=-\frac{1}{2c_{\beta}}\left(m_{H}^{2}c_{\alpha}c_{\beta-\alpha}-m_{h}^{2}s_{\alpha}s_{\beta-\alpha}\right)+m_{A}^{2}s_{\beta}^{2};\\
    \mu_{\eta}^{2} &=-\frac{1}{2s_{\beta}}\left(m_{h}^{2}c_{\alpha}s_{\beta-\alpha}+m_{H}^{2}s_{\alpha}c_{\beta-\alpha}\right)+m_{A}^{2}c_{\beta}^{2};
\end{align}

\section{RGE for the effective 3-3-1 model}
\label{ap:rge}

In view of the expressions presented in the previous appendix, we can easily obtain the quartic couplings $\lambda_1$, $\lambda_2$, $\lambda_{12}$, $\zeta_{12}$ once the masses of the scalars particles and mixing angles are provided. By assuming the limit $v_{\chi}\gg v_{\rho},v_{\eta}$, we consider that the scalar potential of our theory consists of eq.~\ref{n2} while the Yukawa Lagrangian is given by eqs.~\ref{eq:n2HDM}-\ref{eq:n2HDM2}. Thus, the renormalization group equations (RGE) can be straightforward obtained by SARAH \cite{Staub:2013tta}, whose expressions, for the quartic couplings, we reproduce below
{\allowdisplaybreaks  \begin{align} 
16\pi^{2}\beta_{\lambda_1}  =&  
\frac{27}{200} g_{1}^{4} +\frac{9}{20} g_{1}^{2} g_{2}^{2} +\frac{9}{8} g_{2}^{4} -\frac{9}{5} g_{1}^{2} \lambda_1 -9 g_{2}^{2} \lambda_1 +24 \lambda_{1}^{2} +2 \lambda_{12}^{2} +2 \lambda_{12} \zeta_{12} +\zeta_{12}^{2}\\&+12 \lambda_1 y_{t}^{2} -6 y_{t}^{4} \nonumber\\ 
16\pi^{2}\beta_{\lambda_2}  = & 
\frac{27}{200} g_{1}^{4} +\frac{9}{20} g_{1}^{2} g_{2}^{2} +\frac{9}{8} g_{2}^{4} -\frac{9}{5} g_{1}^{2} \lambda_2 -9 g_{2}^{2} \lambda_2 +24 \lambda_{2}^{2} +2 \lambda_{12}^{2} +2 \lambda_{12} \zeta_{12} +\zeta_{12}^{2}\\
16\pi^{2}\beta_{\lambda_{12}}  = & 
\frac{27}{100} g_{1}^{4} +\frac{9}{10} g_{1}^{2} g_{2}^{2} +\frac{9}{4} g_{2}^{4} -\frac{9}{5} g_{1}^{2} \lambda_{12} -9 g_{2}^{2} \lambda_{12} +12 \lambda_1 \lambda_{12} +12 \lambda_2 \lambda_{12} +4 \lambda_{12}^{2} \\&+4 \lambda_1 \zeta_{12} +4 \lambda_2 \zeta_{12} +2 \zeta_{12}^{2} 
 +6 \lambda_{12} y_{t}^{2} \nonumber\\ 
16\pi^{2}\beta_{\zeta_{12}}  = &  
-\frac{9}{5} g_{1}^{2} g_{2}^{2} -\frac{9}{5} g_{1}^{2} \zeta_{12} -9 g_{2}^{2} \zeta_{12} +4 \lambda_1 \zeta_{12} +4 \lambda_2 \zeta_{12} +8 \lambda_{12} \zeta_{12} +4 \zeta_{12}^{2} +6 \zeta_{12} y_{t}^{2} 
\end{align}} 

Regarding the fermions, we only consider the influence of top which give the larger contribution. When performing the running, we used the following SM input values at the EW scale (for definiteness, we will adopt the top mass as the EW scale)\;\cite{Buttazzo:2013uya}
\begin{equation}
 g_{1}=0.3587\,;\quad g_{2} = 0.6483\,;\quad v = 246.924\, \unit{GeV}\,;\quad y_{t} = 0.93697\;.
\end{equation}

\bibliography{main}

\begin{thebibliography}{111}
\expandafter\ifx\csname natexlab\endcsname\relax\def\natexlab#1{#1}\fi
\expandafter\ifx\csname bibnamefont\endcsname\relax
  \def\bibnamefont#1{#1}\fi
\expandafter\ifx\csname bibfnamefont\endcsname\relax
  \def\bibfnamefont#1{#1}\fi
\expandafter\ifx\csname citenamefont\endcsname\relax
  \def\citenamefont#1{#1}\fi
\expandafter\ifx\csname url\endcsname\relax
  \def\url#1{\texttt{#1}}\fi
\expandafter\ifx\csname urlprefix\endcsname\relax\def\urlprefix{URL }\fi
\providecommand{\bibinfo}[2]{#2}
\providecommand{\eprint}[2][]{\url{#2}}

\bibitem[{\citenamefont{Pisano and Pleitez}(1992)}]{Pisano:1992bxx}
\bibinfo{author}{\bibfnamefont{F.}~\bibnamefont{Pisano}} \bibnamefont{and}
  \bibinfo{author}{\bibfnamefont{V.}~\bibnamefont{Pleitez}},
  \bibinfo{journal}{Phys. Rev. D} \textbf{\bibinfo{volume}{46}},
  \bibinfo{pages}{410} (\bibinfo{year}{1992}), \eprint{hep-ph/9206242}.

\bibitem[{\citenamefont{Frampton}(1992)}]{Frampton:1992wt}
\bibinfo{author}{\bibfnamefont{P.~H.} \bibnamefont{Frampton}},
  \bibinfo{journal}{Phys. Rev. Lett.} \textbf{\bibinfo{volume}{69}},
  \bibinfo{pages}{2889} (\bibinfo{year}{1992}).

\bibitem[{\citenamefont{Montero et~al.}(2001)\citenamefont{Montero,
  de~S.~Pires, and Pleitez}}]{Montero:2000rh}
\bibinfo{author}{\bibfnamefont{J.~C.} \bibnamefont{Montero}},
  \bibinfo{author}{\bibfnamefont{C.~A.} \bibnamefont{de~S.~Pires}},
  \bibnamefont{and} \bibinfo{author}{\bibfnamefont{V.}~\bibnamefont{Pleitez}},
  \bibinfo{journal}{Phys. Lett. B} \textbf{\bibinfo{volume}{502}},
  \bibinfo{pages}{167} (\bibinfo{year}{2001}), \eprint{hep-ph/0011296}.

\bibitem[{\citenamefont{Tully and Joshi}(2001)}]{Tully:2000kk}
\bibinfo{author}{\bibfnamefont{M.~B.} \bibnamefont{Tully}} \bibnamefont{and}
  \bibinfo{author}{\bibfnamefont{G.~C.} \bibnamefont{Joshi}},
  \bibinfo{journal}{Phys. Rev. D} \textbf{\bibinfo{volume}{64}},
  \bibinfo{pages}{011301} (\bibinfo{year}{2001}), \eprint{hep-ph/0011172}.

\bibitem[{\citenamefont{Montero et~al.}(2002)\citenamefont{Montero,
  De~S.~Pires, and Pleitez}}]{Montero:2001ts}
\bibinfo{author}{\bibfnamefont{J.~C.} \bibnamefont{Montero}},
  \bibinfo{author}{\bibfnamefont{C.~A.} \bibnamefont{De~S.~Pires}},
  \bibnamefont{and} \bibinfo{author}{\bibfnamefont{V.}~\bibnamefont{Pleitez}},
  \bibinfo{journal}{Phys. Rev. D} \textbf{\bibinfo{volume}{65}},
  \bibinfo{pages}{095001} (\bibinfo{year}{2002}), \eprint{hep-ph/0112246}.

\bibitem[{\citenamefont{Cortez and Tonasse}(2005)}]{Cortez:2005cp}
\bibinfo{author}{\bibfnamefont{N.~V.} \bibnamefont{Cortez}} \bibnamefont{and}
  \bibinfo{author}{\bibfnamefont{M.~D.} \bibnamefont{Tonasse}},
  \bibinfo{journal}{Phys. Rev. D} \textbf{\bibinfo{volume}{72}},
  \bibinfo{pages}{073005} (\bibinfo{year}{2005}), \eprint{hep-ph/0510143}.

\bibitem[{\citenamefont{Cogollo et~al.}(2009)\citenamefont{Cogollo, Diniz, and
  de~S.~Pires}}]{Cogollo:2009yi}
\bibinfo{author}{\bibfnamefont{D.}~\bibnamefont{Cogollo}},
  \bibinfo{author}{\bibfnamefont{H.}~\bibnamefont{Diniz}}, \bibnamefont{and}
  \bibinfo{author}{\bibfnamefont{C.~A.} \bibnamefont{de~S.~Pires}},
  \bibinfo{journal}{Phys. Lett. B} \textbf{\bibinfo{volume}{677}},
  \bibinfo{pages}{338} (\bibinfo{year}{2009}), \eprint{0903.0370}.

\bibitem[{\citenamefont{Cogollo et~al.}(2010)\citenamefont{Cogollo, Diniz, and
  de~S.~Pires}}]{Cogollo:2010jw}
\bibinfo{author}{\bibfnamefont{D.}~\bibnamefont{Cogollo}},
  \bibinfo{author}{\bibfnamefont{H.}~\bibnamefont{Diniz}}, \bibnamefont{and}
  \bibinfo{author}{\bibfnamefont{C.~A.} \bibnamefont{de~S.~Pires}},
  \bibinfo{journal}{Phys. Lett. B} \textbf{\bibinfo{volume}{687}},
  \bibinfo{pages}{400} (\bibinfo{year}{2010}), \eprint{1002.1944}.

\bibitem[{\citenamefont{Cogollo et~al.}(2008)\citenamefont{Cogollo, Diniz,
  de~S.~Pires, and Rodrigues~da Silva}}]{Cogollo:2008zc}
\bibinfo{author}{\bibfnamefont{D.}~\bibnamefont{Cogollo}},
  \bibinfo{author}{\bibfnamefont{H.}~\bibnamefont{Diniz}},
  \bibinfo{author}{\bibfnamefont{C.~A.} \bibnamefont{de~S.~Pires}},
  \bibnamefont{and} \bibinfo{author}{\bibfnamefont{P.~S.}
  \bibnamefont{Rodrigues~da Silva}}, \bibinfo{journal}{Eur. Phys. J. C}
  \textbf{\bibinfo{volume}{58}}, \bibinfo{pages}{455} (\bibinfo{year}{2008}),
  \eprint{0806.3087}.

\bibitem[{\citenamefont{Dias et~al.}(2012)\citenamefont{Dias, de~S.~Pires,
  Rodrigues~da Silva, and Sampieri}}]{Dias:2012xp}
\bibinfo{author}{\bibfnamefont{A.~G.} \bibnamefont{Dias}},
  \bibinfo{author}{\bibfnamefont{C.~A.} \bibnamefont{de~S.~Pires}},
  \bibinfo{author}{\bibfnamefont{P.~S.} \bibnamefont{Rodrigues~da Silva}},
  \bibnamefont{and} \bibinfo{author}{\bibfnamefont{A.}~\bibnamefont{Sampieri}},
  \bibinfo{journal}{Phys. Rev. D} \textbf{\bibinfo{volume}{86}},
  \bibinfo{pages}{035007} (\bibinfo{year}{2012}), \eprint{1206.2590}.

\bibitem[{\citenamefont{Okada et~al.}(2016{\natexlab{a}})\citenamefont{Okada,
  Okada, and Orikasa}}]{Okada:2015bxa}
\bibinfo{author}{\bibfnamefont{H.}~\bibnamefont{Okada}},
  \bibinfo{author}{\bibfnamefont{N.}~\bibnamefont{Okada}}, \bibnamefont{and}
  \bibinfo{author}{\bibfnamefont{Y.}~\bibnamefont{Orikasa}},
  \bibinfo{journal}{Phys. Rev. D} \textbf{\bibinfo{volume}{93}},
  \bibinfo{pages}{073006} (\bibinfo{year}{2016}{\natexlab{a}}),
  \eprint{1504.01204}.

\bibitem[{\citenamefont{Vien et~al.}(2019)\citenamefont{Vien, Long, and
  C\'arcamo~Hern\'andez}}]{Vien:2018otl}
\bibinfo{author}{\bibfnamefont{V.~V.} \bibnamefont{Vien}},
  \bibinfo{author}{\bibfnamefont{H.~N.} \bibnamefont{Long}}, \bibnamefont{and}
  \bibinfo{author}{\bibfnamefont{A.~E.} \bibnamefont{C\'arcamo~Hern\'andez}},
  \bibinfo{journal}{Mod. Phys. Lett. A} \textbf{\bibinfo{volume}{34}},
  \bibinfo{pages}{1950005} (\bibinfo{year}{2019}), \eprint{1812.07263}.

\bibitem[{\citenamefont{C\'arcamo~Hern\'andez
  et~al.}(2018)\citenamefont{C\'arcamo~Hern\'andez, Long, and
  Vien}}]{carcamoHernandez:2018iel}
\bibinfo{author}{\bibfnamefont{A.~E.} \bibnamefont{C\'arcamo~Hern\'andez}},
  \bibinfo{author}{\bibfnamefont{H.~N.} \bibnamefont{Long}}, \bibnamefont{and}
  \bibinfo{author}{\bibfnamefont{V.~V.} \bibnamefont{Vien}},
  \bibinfo{journal}{Eur. Phys. J. C} \textbf{\bibinfo{volume}{78}},
  \bibinfo{pages}{804} (\bibinfo{year}{2018}), \eprint{1803.01636}.

\bibitem[{\citenamefont{Nguyen et~al.}(2018)\citenamefont{Nguyen, Le, Hong, and
  Hue}}]{Nguyen:2018rlb}
\bibinfo{author}{\bibfnamefont{T.~P.} \bibnamefont{Nguyen}},
  \bibinfo{author}{\bibfnamefont{T.~T.} \bibnamefont{Le}},
  \bibinfo{author}{\bibfnamefont{T.~T.} \bibnamefont{Hong}}, \bibnamefont{and}
  \bibinfo{author}{\bibfnamefont{L.~T.} \bibnamefont{Hue}},
  \bibinfo{journal}{Phys. Rev. D} \textbf{\bibinfo{volume}{97}},
  \bibinfo{pages}{073003} (\bibinfo{year}{2018}), \eprint{1802.00429}.

\bibitem[{\citenamefont{de~Sousa~Pires
  et~al.}(2019)\citenamefont{de~Sousa~Pires, Ferreira De~Freitas, Shu, Huang,
  and Wagner Vasconcelos~Oleg\'ario}}]{Pires:2018kaj}
\bibinfo{author}{\bibfnamefont{C.~A.} \bibnamefont{de~Sousa~Pires}},
  \bibinfo{author}{\bibfnamefont{F.}~\bibnamefont{Ferreira De~Freitas}},
  \bibinfo{author}{\bibfnamefont{J.}~\bibnamefont{Shu}},
  \bibinfo{author}{\bibfnamefont{L.}~\bibnamefont{Huang}}, \bibnamefont{and}
  \bibinfo{author}{\bibfnamefont{P.}~\bibnamefont{Wagner
  Vasconcelos~Oleg\'ario}}, \bibinfo{journal}{Phys. Lett. B}
  \textbf{\bibinfo{volume}{797}}, \bibinfo{pages}{134827}
  (\bibinfo{year}{2019}), \eprint{1812.10570}.

\bibitem[{\citenamefont{C\'arcamo~Hern\'andez
  et~al.}(2019{\natexlab{a}})\citenamefont{C\'arcamo~Hern\'andez,
  P\'erez-Julve, and Hidalgo~Vel\'asquez}}]{CarcamoHernandez:2019iwh}
\bibinfo{author}{\bibfnamefont{A.~E.} \bibnamefont{C\'arcamo~Hern\'andez}},
  \bibinfo{author}{\bibfnamefont{N.~A.} \bibnamefont{P\'erez-Julve}},
  \bibnamefont{and}
  \bibinfo{author}{\bibfnamefont{Y.}~\bibnamefont{Hidalgo~Vel\'asquez}},
  \bibinfo{journal}{Phys. Rev. D} \textbf{\bibinfo{volume}{100}},
  \bibinfo{pages}{095025} (\bibinfo{year}{2019}{\natexlab{a}}),
  \eprint{1907.13083}.

\bibitem[{\citenamefont{C\'arcamo~Hern\'andez
  et~al.}(2019{\natexlab{b}})\citenamefont{C\'arcamo~Hern\'andez,
  Hidalgo~Vel\'asquez, and P\'erez-Julve}}]{CarcamoHernandez:2019vih}
\bibinfo{author}{\bibfnamefont{A.~E.} \bibnamefont{C\'arcamo~Hern\'andez}},
  \bibinfo{author}{\bibfnamefont{Y.}~\bibnamefont{Hidalgo~Vel\'asquez}},
  \bibnamefont{and} \bibinfo{author}{\bibfnamefont{N.~A.}
  \bibnamefont{P\'erez-Julve}}, \bibinfo{journal}{Eur. Phys. J. C}
  \textbf{\bibinfo{volume}{79}}, \bibinfo{pages}{828}
  (\bibinfo{year}{2019}{\natexlab{b}}), \eprint{1905.02323}.

\bibitem[{\citenamefont{C\'arcamo~Hern\'andez
  et~al.}(2021)\citenamefont{C\'arcamo~Hern\'andez, Hue, Kovalenko, and
  Long}}]{CarcamoHernandez:2020pnh}
\bibinfo{author}{\bibfnamefont{A.~E.} \bibnamefont{C\'arcamo~Hern\'andez}},
  \bibinfo{author}{\bibfnamefont{L.~T.} \bibnamefont{Hue}},
  \bibinfo{author}{\bibfnamefont{S.}~\bibnamefont{Kovalenko}},
  \bibnamefont{and} \bibinfo{author}{\bibfnamefont{H.~N.} \bibnamefont{Long}},
  \bibinfo{journal}{Eur. Phys. J. Plus} \textbf{\bibinfo{volume}{136}},
  \bibinfo{pages}{1158} (\bibinfo{year}{2021}), \eprint{2001.01748}.

\bibitem[{\citenamefont{Fregolente and Tonasse}(2003)}]{Fregolente:2002nx}
\bibinfo{author}{\bibfnamefont{D.}~\bibnamefont{Fregolente}} \bibnamefont{and}
  \bibinfo{author}{\bibfnamefont{M.~D.} \bibnamefont{Tonasse}},
  \bibinfo{journal}{Phys. Lett. B} \textbf{\bibinfo{volume}{555}},
  \bibinfo{pages}{7} (\bibinfo{year}{2003}), \eprint{hep-ph/0209119}.

\bibitem[{\citenamefont{Long and Lan}(2003)}]{Hoang:2003vj}
\bibinfo{author}{\bibfnamefont{H.~N.} \bibnamefont{Long}} \bibnamefont{and}
  \bibinfo{author}{\bibfnamefont{N.~Q.} \bibnamefont{Lan}},
  \bibinfo{journal}{EPL} \textbf{\bibinfo{volume}{64}}, \bibinfo{pages}{571}
  (\bibinfo{year}{2003}), \eprint{hep-ph/0309038}.

\bibitem[{\citenamefont{de~S.~Pires and Rodrigues~da
  Silva}(2007)}]{deS.Pires:2007gi}
\bibinfo{author}{\bibfnamefont{C.~A.} \bibnamefont{de~S.~Pires}}
  \bibnamefont{and} \bibinfo{author}{\bibfnamefont{P.~S.}
  \bibnamefont{Rodrigues~da Silva}}, \bibinfo{journal}{JCAP}
  \textbf{\bibinfo{volume}{12}}, \bibinfo{pages}{012} (\bibinfo{year}{2007}),
  \eprint{0710.2104}.

\bibitem[{\citenamefont{Mizukoshi et~al.}(2011)\citenamefont{Mizukoshi,
  de~S.~Pires, Queiroz, and Rodrigues~da Silva}}]{Mizukoshi:2010ky}
\bibinfo{author}{\bibfnamefont{J.~K.} \bibnamefont{Mizukoshi}},
  \bibinfo{author}{\bibfnamefont{C.~A.} \bibnamefont{de~S.~Pires}},
  \bibinfo{author}{\bibfnamefont{F.~S.} \bibnamefont{Queiroz}},
  \bibnamefont{and} \bibinfo{author}{\bibfnamefont{P.~S.}
  \bibnamefont{Rodrigues~da Silva}}, \bibinfo{journal}{Phys. Rev. D}
  \textbf{\bibinfo{volume}{83}}, \bibinfo{pages}{065024}
  (\bibinfo{year}{2011}), \eprint{1010.4097}.

\bibitem[{\citenamefont{Ruiz-Alvarez et~al.}(2012)\citenamefont{Ruiz-Alvarez,
  de~S.~Pires, Queiroz, Restrepo, and Rodrigues~da
  Silva}}]{Ruiz-Alvarez:2012nvg}
\bibinfo{author}{\bibfnamefont{J.~D.} \bibnamefont{Ruiz-Alvarez}},
  \bibinfo{author}{\bibfnamefont{C.~A.} \bibnamefont{de~S.~Pires}},
  \bibinfo{author}{\bibfnamefont{F.~S.} \bibnamefont{Queiroz}},
  \bibinfo{author}{\bibfnamefont{D.}~\bibnamefont{Restrepo}}, \bibnamefont{and}
  \bibinfo{author}{\bibfnamefont{P.~S.} \bibnamefont{Rodrigues~da Silva}},
  \bibinfo{journal}{Phys. Rev. D} \textbf{\bibinfo{volume}{86}},
  \bibinfo{pages}{075011} (\bibinfo{year}{2012}), \eprint{1206.5779}.

\bibitem[{\citenamefont{Profumo and Queiroz}(2014)}]{Profumo:2013sca}
\bibinfo{author}{\bibfnamefont{S.}~\bibnamefont{Profumo}} \bibnamefont{and}
  \bibinfo{author}{\bibfnamefont{F.~S.} \bibnamefont{Queiroz}},
  \bibinfo{journal}{Eur. Phys. J. C} \textbf{\bibinfo{volume}{74}},
  \bibinfo{pages}{2960} (\bibinfo{year}{2014}), \eprint{1307.7802}.

\bibitem[{\citenamefont{Dong et~al.}(2013{\natexlab{a}})\citenamefont{Dong,
  Nguyen, and Soa}}]{Dong:2013ioa}
\bibinfo{author}{\bibfnamefont{P.~V.} \bibnamefont{Dong}},
  \bibinfo{author}{\bibfnamefont{T.~P.} \bibnamefont{Nguyen}},
  \bibnamefont{and} \bibinfo{author}{\bibfnamefont{D.~V.} \bibnamefont{Soa}},
  \bibinfo{journal}{Phys. Rev. D} \textbf{\bibinfo{volume}{88}},
  \bibinfo{pages}{095014} (\bibinfo{year}{2013}{\natexlab{a}}),
  \eprint{1308.4097}.

\bibitem[{\citenamefont{Dong et~al.}(2013{\natexlab{b}})\citenamefont{Dong,
  Hung, and Tham}}]{Dong:2013wca}
\bibinfo{author}{\bibfnamefont{P.~V.} \bibnamefont{Dong}},
  \bibinfo{author}{\bibfnamefont{H.~T.} \bibnamefont{Hung}}, \bibnamefont{and}
  \bibinfo{author}{\bibfnamefont{T.~D.} \bibnamefont{Tham}},
  \bibinfo{journal}{Phys. Rev. D} \textbf{\bibinfo{volume}{87}},
  \bibinfo{pages}{115003} (\bibinfo{year}{2013}{\natexlab{b}}),
  \eprint{1305.0369}.

\bibitem[{\citenamefont{Cogollo
  et~al.}(2014{\natexlab{a}})\citenamefont{Cogollo, Gonzalez-Morales, Queiroz,
  and Teles}}]{Cogollo:2014jia}
\bibinfo{author}{\bibfnamefont{D.}~\bibnamefont{Cogollo}},
  \bibinfo{author}{\bibfnamefont{A.~X.} \bibnamefont{Gonzalez-Morales}},
  \bibinfo{author}{\bibfnamefont{F.~S.} \bibnamefont{Queiroz}},
  \bibnamefont{and} \bibinfo{author}{\bibfnamefont{P.~R.} \bibnamefont{Teles}},
  \bibinfo{journal}{JCAP} \textbf{\bibinfo{volume}{11}}, \bibinfo{pages}{002}
  (\bibinfo{year}{2014}{\natexlab{a}}), \eprint{1402.3271}.

\bibitem[{\citenamefont{Dong et~al.}(2014{\natexlab{a}})\citenamefont{Dong,
  Huong, Queiroz, and Thuy}}]{Dong:2014wsa}
\bibinfo{author}{\bibfnamefont{P.~V.} \bibnamefont{Dong}},
  \bibinfo{author}{\bibfnamefont{D.~T.} \bibnamefont{Huong}},
  \bibinfo{author}{\bibfnamefont{F.~S.} \bibnamefont{Queiroz}},
  \bibnamefont{and} \bibinfo{author}{\bibfnamefont{N.~T.} \bibnamefont{Thuy}},
  \bibinfo{journal}{Phys. Rev. D} \textbf{\bibinfo{volume}{90}},
  \bibinfo{pages}{075021} (\bibinfo{year}{2014}{\natexlab{a}}),
  \eprint{1405.2591}.

\bibitem[{\citenamefont{Dong et~al.}(2014{\natexlab{b}})\citenamefont{Dong,
  Ngan, and Soa}}]{Dong:2014esa}
\bibinfo{author}{\bibfnamefont{P.~V.} \bibnamefont{Dong}},
  \bibinfo{author}{\bibfnamefont{N.~T.~K.} \bibnamefont{Ngan}},
  \bibnamefont{and} \bibinfo{author}{\bibfnamefont{D.~V.} \bibnamefont{Soa}},
  \bibinfo{journal}{Phys. Rev. D} \textbf{\bibinfo{volume}{90}},
  \bibinfo{pages}{075019} (\bibinfo{year}{2014}{\natexlab{b}}),
  \eprint{1407.3839}.

\bibitem[{\citenamefont{Kelso et~al.}(2014)\citenamefont{Kelso, Long, Martinez,
  and Queiroz}}]{Kelso:2014qka}
\bibinfo{author}{\bibfnamefont{C.}~\bibnamefont{Kelso}},
  \bibinfo{author}{\bibfnamefont{H.~N.} \bibnamefont{Long}},
  \bibinfo{author}{\bibfnamefont{R.}~\bibnamefont{Martinez}}, \bibnamefont{and}
  \bibinfo{author}{\bibfnamefont{F.~S.} \bibnamefont{Queiroz}},
  \bibinfo{journal}{Phys. Rev. D} \textbf{\bibinfo{volume}{90}},
  \bibinfo{pages}{113011} (\bibinfo{year}{2014}), \eprint{1408.6203}.

\bibitem[{\citenamefont{Mambrini et~al.}(2016)\citenamefont{Mambrini, Profumo,
  and Queiroz}}]{Mambrini:2015sia}
\bibinfo{author}{\bibfnamefont{Y.}~\bibnamefont{Mambrini}},
  \bibinfo{author}{\bibfnamefont{S.}~\bibnamefont{Profumo}}, \bibnamefont{and}
  \bibinfo{author}{\bibfnamefont{F.~S.} \bibnamefont{Queiroz}},
  \bibinfo{journal}{Phys. Lett. B} \textbf{\bibinfo{volume}{760}},
  \bibinfo{pages}{807} (\bibinfo{year}{2016}), \eprint{1508.06635}.

\bibitem[{\citenamefont{Dong et~al.}(2015)\citenamefont{Dong, Kim, Soa, and
  Thuy}}]{Dong:2015rka}
\bibinfo{author}{\bibfnamefont{P.~V.} \bibnamefont{Dong}},
  \bibinfo{author}{\bibfnamefont{C.~S.} \bibnamefont{Kim}},
  \bibinfo{author}{\bibfnamefont{D.~V.} \bibnamefont{Soa}}, \bibnamefont{and}
  \bibinfo{author}{\bibfnamefont{N.~T.} \bibnamefont{Thuy}},
  \bibinfo{journal}{Phys. Rev. D} \textbf{\bibinfo{volume}{91}},
  \bibinfo{pages}{115019} (\bibinfo{year}{2015}), \eprint{1501.04385}.

\bibitem[{\citenamefont{de~S.~Pires et~al.}(2016)\citenamefont{de~S.~Pires,
  Rodrigues~da Silva, Santos, and Siqueira}}]{deSPires:2016kkg}
\bibinfo{author}{\bibfnamefont{C.~A.} \bibnamefont{de~S.~Pires}},
  \bibinfo{author}{\bibfnamefont{P.~S.} \bibnamefont{Rodrigues~da Silva}},
  \bibinfo{author}{\bibfnamefont{A.~C.~O.} \bibnamefont{Santos}},
  \bibnamefont{and} \bibinfo{author}{\bibfnamefont{C.}~\bibnamefont{Siqueira}},
  \bibinfo{journal}{Phys. Rev. D} \textbf{\bibinfo{volume}{94}},
  \bibinfo{pages}{055014} (\bibinfo{year}{2016}), \eprint{1606.01853}.

\bibitem[{\citenamefont{Alves et~al.}(2017)\citenamefont{Alves, Arcadi, Dong,
  Duarte, Queiroz, and Valle}}]{Alves:2016fqe}
\bibinfo{author}{\bibfnamefont{A.}~\bibnamefont{Alves}},
  \bibinfo{author}{\bibfnamefont{G.}~\bibnamefont{Arcadi}},
  \bibinfo{author}{\bibfnamefont{P.~V.} \bibnamefont{Dong}},
  \bibinfo{author}{\bibfnamefont{L.}~\bibnamefont{Duarte}},
  \bibinfo{author}{\bibfnamefont{F.~S.} \bibnamefont{Queiroz}},
  \bibnamefont{and} \bibinfo{author}{\bibfnamefont{J.~W.~F.}
  \bibnamefont{Valle}}, \bibinfo{journal}{Phys. Lett. B}
  \textbf{\bibinfo{volume}{772}}, \bibinfo{pages}{825} (\bibinfo{year}{2017}),
  \eprint{1612.04383}.

\bibitem[{\citenamefont{Rodrigues~da Silva}(2016)}]{RodriguesdaSilva:2014gbi}
\bibinfo{author}{\bibfnamefont{P.~S.} \bibnamefont{Rodrigues~da Silva}},
  \bibinfo{journal}{Phys. Int.} \textbf{\bibinfo{volume}{7}},
  \bibinfo{pages}{15} (\bibinfo{year}{2016}), \eprint{1412.8633}.

\bibitem[{\citenamefont{Carvajal et~al.}(2017)\citenamefont{Carvajal,
  S\'anchez-Vega, and Zapata}}]{Carvajal:2017gjj}
\bibinfo{author}{\bibfnamefont{C.~D.~R.} \bibnamefont{Carvajal}},
  \bibinfo{author}{\bibfnamefont{B.~L.} \bibnamefont{S\'anchez-Vega}},
  \bibnamefont{and} \bibinfo{author}{\bibfnamefont{O.}~\bibnamefont{Zapata}},
  \bibinfo{journal}{Phys. Rev. D} \textbf{\bibinfo{volume}{96}},
  \bibinfo{pages}{115035} (\bibinfo{year}{2017}), \eprint{1704.08340}.

\bibitem[{\citenamefont{Dong et~al.}(2018)\citenamefont{Dong, Huong, Queiroz,
  Valle, and Vaquera-Araujo}}]{Dong:2017zxo}
\bibinfo{author}{\bibfnamefont{P.~V.} \bibnamefont{Dong}},
  \bibinfo{author}{\bibfnamefont{D.~T.} \bibnamefont{Huong}},
  \bibinfo{author}{\bibfnamefont{F.~S.} \bibnamefont{Queiroz}},
  \bibinfo{author}{\bibfnamefont{J.~W.~F.} \bibnamefont{Valle}},
  \bibnamefont{and} \bibinfo{author}{\bibfnamefont{C.~A.}
  \bibnamefont{Vaquera-Araujo}}, \bibinfo{journal}{JHEP}
  \textbf{\bibinfo{volume}{04}}, \bibinfo{pages}{143} (\bibinfo{year}{2018}),
  \eprint{1710.06951}.

\bibitem[{\citenamefont{Arcadi et~al.}(2018)\citenamefont{Arcadi, Ferreira,
  Goertz, Guzzo, Queiroz, and Santos}}]{Arcadi:2017xbo}
\bibinfo{author}{\bibfnamefont{G.}~\bibnamefont{Arcadi}},
  \bibinfo{author}{\bibfnamefont{C.~P.} \bibnamefont{Ferreira}},
  \bibinfo{author}{\bibfnamefont{F.}~\bibnamefont{Goertz}},
  \bibinfo{author}{\bibfnamefont{M.~M.} \bibnamefont{Guzzo}},
  \bibinfo{author}{\bibfnamefont{F.~S.} \bibnamefont{Queiroz}},
  \bibnamefont{and} \bibinfo{author}{\bibfnamefont{A.~C.~O.}
  \bibnamefont{Santos}}, \bibinfo{journal}{Phys. Rev. D}
  \textbf{\bibinfo{volume}{97}}, \bibinfo{pages}{075022}
  (\bibinfo{year}{2018}), \eprint{1712.02373}.

\bibitem[{\citenamefont{Montero et~al.}(2018)\citenamefont{Montero, Romero, and
  S\'anchez-Vega}}]{Montero:2017yvy}
\bibinfo{author}{\bibfnamefont{J.~C.} \bibnamefont{Montero}},
  \bibinfo{author}{\bibfnamefont{A.}~\bibnamefont{Romero}}, \bibnamefont{and}
  \bibinfo{author}{\bibfnamefont{B.~L.} \bibnamefont{S\'anchez-Vega}},
  \bibinfo{journal}{Phys. Rev. D} \textbf{\bibinfo{volume}{97}},
  \bibinfo{pages}{063015} (\bibinfo{year}{2018}), \eprint{1709.04535}.

\bibitem[{\citenamefont{Huong et~al.}(2019)\citenamefont{Huong, Dinh, Thien,
  and Van~Dong}}]{Huong:2019vej}
\bibinfo{author}{\bibfnamefont{D.~T.} \bibnamefont{Huong}},
  \bibinfo{author}{\bibfnamefont{D.~N.} \bibnamefont{Dinh}},
  \bibinfo{author}{\bibfnamefont{L.~D.} \bibnamefont{Thien}}, \bibnamefont{and}
  \bibinfo{author}{\bibfnamefont{P.}~\bibnamefont{Van~Dong}},
  \bibinfo{journal}{JHEP} \textbf{\bibinfo{volume}{08}}, \bibinfo{pages}{051}
  (\bibinfo{year}{2019}), \eprint{1906.05240}.

\bibitem[{\citenamefont{Alvarez-Salazar and
  Peres}(2021)}]{Alvarez-Salazar:2019cxw}
\bibinfo{author}{\bibfnamefont{C.~E.} \bibnamefont{Alvarez-Salazar}}
  \bibnamefont{and} \bibinfo{author}{\bibfnamefont{O.~L.~G.}
  \bibnamefont{Peres}}, \bibinfo{journal}{Phys. Rev. D}
  \textbf{\bibinfo{volume}{103}}, \bibinfo{pages}{035029}
  (\bibinfo{year}{2021}), \eprint{1906.06444}.

\bibitem[{\citenamefont{Van~Loi et~al.}(2021)\citenamefont{Van~Loi, Nam, and
  Van~Dong}}]{VanLoi:2020xcq}
\bibinfo{author}{\bibfnamefont{D.}~\bibnamefont{Van~Loi}},
  \bibinfo{author}{\bibfnamefont{C.~H.} \bibnamefont{Nam}}, \bibnamefont{and}
  \bibinfo{author}{\bibfnamefont{P.}~\bibnamefont{Van~Dong}},
  \bibinfo{journal}{Eur. Phys. J. C} \textbf{\bibinfo{volume}{81}},
  \bibinfo{pages}{591} (\bibinfo{year}{2021}), \eprint{2012.10979}.

\bibitem[{\citenamefont{Dutra et~al.}(2021)\citenamefont{Dutra, Oliveira,
  de~S.~Pires, and Queiroz}}]{Dutra:2021lto}
\bibinfo{author}{\bibfnamefont{M.}~\bibnamefont{Dutra}},
  \bibinfo{author}{\bibfnamefont{V.}~\bibnamefont{Oliveira}},
  \bibinfo{author}{\bibfnamefont{C.~A.} \bibnamefont{de~S.~Pires}},
  \bibnamefont{and} \bibinfo{author}{\bibfnamefont{F.~S.}
  \bibnamefont{Queiroz}}, \bibinfo{journal}{JHEP}
  \textbf{\bibinfo{volume}{10}}, \bibinfo{pages}{005} (\bibinfo{year}{2021}),
  \eprint{2104.14542}.

\bibitem[{\citenamefont{Oliveira and de~S.~Pires}(2022)}]{Oliveira:2021gcw}
\bibinfo{author}{\bibfnamefont{V.}~\bibnamefont{Oliveira}} \bibnamefont{and}
  \bibinfo{author}{\bibfnamefont{C.~A.} \bibnamefont{de~S.~Pires}},
  \bibinfo{journal}{Phys. Rev. D} \textbf{\bibinfo{volume}{106}},
  \bibinfo{pages}{015031} (\bibinfo{year}{2022}), \eprint{2112.03963}.

\bibitem[{\citenamefont{Cogollo et~al.}(2012)\citenamefont{Cogollo, de~Andrade,
  Queiroz, and Rebello~Teles}}]{Cogollo:2012ek}
\bibinfo{author}{\bibfnamefont{D.}~\bibnamefont{Cogollo}},
  \bibinfo{author}{\bibfnamefont{A.~V.} \bibnamefont{de~Andrade}},
  \bibinfo{author}{\bibfnamefont{F.~S.} \bibnamefont{Queiroz}},
  \bibnamefont{and}
  \bibinfo{author}{\bibfnamefont{P.}~\bibnamefont{Rebello~Teles}},
  \bibinfo{journal}{Eur. Phys. J. C} \textbf{\bibinfo{volume}{72}},
  \bibinfo{pages}{2029} (\bibinfo{year}{2012}), \eprint{1201.1268}.

\bibitem[{\citenamefont{Cogollo
  et~al.}(2014{\natexlab{b}})\citenamefont{Cogollo, Queiroz, and
  Vasconcelos}}]{Cogollo:2013mga}
\bibinfo{author}{\bibfnamefont{D.}~\bibnamefont{Cogollo}},
  \bibinfo{author}{\bibfnamefont{F.~S.} \bibnamefont{Queiroz}},
  \bibnamefont{and}
  \bibinfo{author}{\bibfnamefont{P.}~\bibnamefont{Vasconcelos}},
  \bibinfo{journal}{Mod. Phys. Lett. A} \textbf{\bibinfo{volume}{29}},
  \bibinfo{pages}{1450173} (\bibinfo{year}{2014}{\natexlab{b}}),
  \eprint{1312.0304}.

\bibitem[{\citenamefont{Buras et~al.}(2014)\citenamefont{Buras, De~Fazio, and
  Girrbach-Noe}}]{Buras:2014yna}
\bibinfo{author}{\bibfnamefont{A.~J.} \bibnamefont{Buras}},
  \bibinfo{author}{\bibfnamefont{F.}~\bibnamefont{De~Fazio}}, \bibnamefont{and}
  \bibinfo{author}{\bibfnamefont{J.}~\bibnamefont{Girrbach-Noe}},
  \bibinfo{journal}{JHEP} \textbf{\bibinfo{volume}{08}}, \bibinfo{pages}{039}
  (\bibinfo{year}{2014}), \eprint{1405.3850}.

\bibitem[{\citenamefont{Buras and
  De~Fazio}(2016{\natexlab{a}})}]{Buras:2015kwd}
\bibinfo{author}{\bibfnamefont{A.~J.} \bibnamefont{Buras}} \bibnamefont{and}
  \bibinfo{author}{\bibfnamefont{F.}~\bibnamefont{De~Fazio}},
  \bibinfo{journal}{JHEP} \textbf{\bibinfo{volume}{03}}, \bibinfo{pages}{010}
  (\bibinfo{year}{2016}{\natexlab{a}}), \eprint{1512.02869}.

\bibitem[{\citenamefont{Queiroz et~al.}(2016)\citenamefont{Queiroz, Siqueira,
  and Valle}}]{Queiroz:2016gif}
\bibinfo{author}{\bibfnamefont{F.~S.} \bibnamefont{Queiroz}},
  \bibinfo{author}{\bibfnamefont{C.}~\bibnamefont{Siqueira}}, \bibnamefont{and}
  \bibinfo{author}{\bibfnamefont{J.~W.~F.} \bibnamefont{Valle}},
  \bibinfo{journal}{Phys. Lett. B} \textbf{\bibinfo{volume}{763}},
  \bibinfo{pages}{269} (\bibinfo{year}{2016}), \eprint{1608.07295}.

\bibitem[{\citenamefont{de~Melo et~al.}(2021)\citenamefont{de~Melo, Kovalenko,
  Queiroz, Siqueira, and Villamizar}}]{deMelo:2021ers}
\bibinfo{author}{\bibfnamefont{T.~B.} \bibnamefont{de~Melo}},
  \bibinfo{author}{\bibfnamefont{S.}~\bibnamefont{Kovalenko}},
  \bibinfo{author}{\bibfnamefont{F.~S.} \bibnamefont{Queiroz}},
  \bibinfo{author}{\bibfnamefont{C.}~\bibnamefont{Siqueira}}, \bibnamefont{and}
  \bibinfo{author}{\bibfnamefont{Y.~S.} \bibnamefont{Villamizar}},
  \bibinfo{journal}{Phys. Rev. D} \textbf{\bibinfo{volume}{103}},
  \bibinfo{pages}{115001} (\bibinfo{year}{2021}), \eprint{2102.06262}.

\bibitem[{\citenamefont{Buras and
  De~Fazio}(2016{\natexlab{b}})}]{Buras:2016dxz}
\bibinfo{author}{\bibfnamefont{A.~J.} \bibnamefont{Buras}} \bibnamefont{and}
  \bibinfo{author}{\bibfnamefont{F.}~\bibnamefont{De~Fazio}},
  \bibinfo{journal}{JHEP} \textbf{\bibinfo{volume}{08}}, \bibinfo{pages}{115}
  (\bibinfo{year}{2016}{\natexlab{b}}), \eprint{1604.02344}.

\bibitem[{\citenamefont{Buras et~al.}(2021)\citenamefont{Buras, Colangelo,
  De~Fazio, and Loparco}}]{Buras:2021rdg}
\bibinfo{author}{\bibfnamefont{A.~J.} \bibnamefont{Buras}},
  \bibinfo{author}{\bibfnamefont{P.}~\bibnamefont{Colangelo}},
  \bibinfo{author}{\bibfnamefont{F.}~\bibnamefont{De~Fazio}}, \bibnamefont{and}
  \bibinfo{author}{\bibfnamefont{F.}~\bibnamefont{Loparco}},
  \bibinfo{journal}{JHEP} \textbf{\bibinfo{volume}{10}}, \bibinfo{pages}{021}
  (\bibinfo{year}{2021}), \eprint{2107.10866}.

\bibitem[{\citenamefont{Pal}(1995)}]{Pal:1994ba}
\bibinfo{author}{\bibfnamefont{P.~B.} \bibnamefont{Pal}},
  \bibinfo{journal}{Phys. Rev. D} \textbf{\bibinfo{volume}{52}},
  \bibinfo{pages}{1659} (\bibinfo{year}{1995}), \eprint{hep-ph/9411406}.

\bibitem[{\citenamefont{Dias and Pleitez}(2004)}]{Dias:2003zt}
\bibinfo{author}{\bibfnamefont{A.~G.} \bibnamefont{Dias}} \bibnamefont{and}
  \bibinfo{author}{\bibfnamefont{V.}~\bibnamefont{Pleitez}},
  \bibinfo{journal}{Phys. Rev. D} \textbf{\bibinfo{volume}{69}},
  \bibinfo{pages}{077702} (\bibinfo{year}{2004}), \eprint{hep-ph/0308037}.

\bibitem[{\citenamefont{Dias et~al.}(2003)\citenamefont{Dias, de~S.~Pires, and
  Rodrigues~da Silva}}]{Dias:2003iq}
\bibinfo{author}{\bibfnamefont{A.~G.} \bibnamefont{Dias}},
  \bibinfo{author}{\bibfnamefont{C.~A.} \bibnamefont{de~S.~Pires}},
  \bibnamefont{and} \bibinfo{author}{\bibfnamefont{P.~S.}
  \bibnamefont{Rodrigues~da Silva}}, \bibinfo{journal}{Phys. Rev. D}
  \textbf{\bibinfo{volume}{68}}, \bibinfo{pages}{115009}
  (\bibinfo{year}{2003}), \eprint{hep-ph/0309058}.

\bibitem[{\citenamefont{Montero and Sanchez-Vega}(2011)}]{Montero:2011tg}
\bibinfo{author}{\bibfnamefont{J.~C.} \bibnamefont{Montero}} \bibnamefont{and}
  \bibinfo{author}{\bibfnamefont{B.~L.} \bibnamefont{Sanchez-Vega}},
  \bibinfo{journal}{Phys. Rev. D} \textbf{\bibinfo{volume}{84}},
  \bibinfo{pages}{055019} (\bibinfo{year}{2011}), \eprint{1102.5374}.

\bibitem[{\citenamefont{Dias et~al.}(2018)\citenamefont{Dias, Leite, Lopes, and
  Nishi}}]{Dias:2018ddy}
\bibinfo{author}{\bibfnamefont{A.~G.} \bibnamefont{Dias}},
  \bibinfo{author}{\bibfnamefont{J.}~\bibnamefont{Leite}},
  \bibinfo{author}{\bibfnamefont{D.~D.} \bibnamefont{Lopes}}, \bibnamefont{and}
  \bibinfo{author}{\bibfnamefont{C.~C.} \bibnamefont{Nishi}},
  \bibinfo{journal}{Phys. Rev. D} \textbf{\bibinfo{volume}{98}},
  \bibinfo{pages}{115017} (\bibinfo{year}{2018}), \eprint{1810.01893}.

\bibitem[{\citenamefont{Dias et~al.}(2020)\citenamefont{Dias, Leite, Valle, and
  Vaquera-Araujo}}]{Dias:2020kbj}
\bibinfo{author}{\bibfnamefont{A.~G.} \bibnamefont{Dias}},
  \bibinfo{author}{\bibfnamefont{J.}~\bibnamefont{Leite}},
  \bibinfo{author}{\bibfnamefont{J.~W.~F.} \bibnamefont{Valle}},
  \bibnamefont{and} \bibinfo{author}{\bibfnamefont{C.~A.}
  \bibnamefont{Vaquera-Araujo}}, \bibinfo{journal}{Phys. Lett. B}
  \textbf{\bibinfo{volume}{810}}, \bibinfo{pages}{135829}
  (\bibinfo{year}{2020}), \eprint{2008.10650}.

\bibitem[{\citenamefont{Alves et~al.}(2022)\citenamefont{Alves, Duarte,
  Kovalenko, Oviedo-Torres, Queiroz, and Villamizar}}]{Alves:2022hcp}
\bibinfo{author}{\bibfnamefont{A.}~\bibnamefont{Alves}},
  \bibinfo{author}{\bibfnamefont{L.}~\bibnamefont{Duarte}},
  \bibinfo{author}{\bibfnamefont{S.}~\bibnamefont{Kovalenko}},
  \bibinfo{author}{\bibfnamefont{Y.~M.} \bibnamefont{Oviedo-Torres}},
  \bibinfo{author}{\bibfnamefont{F.~S.} \bibnamefont{Queiroz}},
  \bibnamefont{and} \bibinfo{author}{\bibfnamefont{Y.~S.}
  \bibnamefont{Villamizar}}, \bibinfo{journal}{Phys. Rev. D}
  \textbf{\bibinfo{volume}{106}}, \bibinfo{pages}{055027}
  (\bibinfo{year}{2022}), \eprint{2203.02520}.

\bibitem[{\citenamefont{Abi et~al.}(2021)}]{Muong-2:2021ojo}
\bibinfo{author}{\bibfnamefont{B.}~\bibnamefont{Abi}} \bibnamefont{et~al.}
  (\bibinfo{collaboration}{Muon g-2}), \bibinfo{journal}{Phys. Rev. Lett.}
  \textbf{\bibinfo{volume}{126}}, \bibinfo{pages}{141801}
  (\bibinfo{year}{2021}), \eprint{2104.03281}.

\bibitem[{\citenamefont{de~Jesus et~al.}(2020)\citenamefont{de~Jesus,
  Kovalenko, Queiroz, de~S.~Pires, and Villamizar}}]{deJesus:2020ngn}
\bibinfo{author}{\bibfnamefont{A.~S.} \bibnamefont{de~Jesus}},
  \bibinfo{author}{\bibfnamefont{S.}~\bibnamefont{Kovalenko}},
  \bibinfo{author}{\bibfnamefont{F.~S.} \bibnamefont{Queiroz}},
  \bibinfo{author}{\bibfnamefont{C.~A.} \bibnamefont{de~S.~Pires}},
  \bibnamefont{and} \bibinfo{author}{\bibfnamefont{Y.~S.}
  \bibnamefont{Villamizar}}, \bibinfo{journal}{Phys. Lett. B}
  \textbf{\bibinfo{volume}{809}}, \bibinfo{pages}{135689}
  (\bibinfo{year}{2020}), \eprint{2003.06440}.

\bibitem[{\citenamefont{Broggio et~al.}(2014)\citenamefont{Broggio, Chun,
  Passera, Patel, and Vempati}}]{Broggio:2014mna}
\bibinfo{author}{\bibfnamefont{A.}~\bibnamefont{Broggio}},
  \bibinfo{author}{\bibfnamefont{E.~J.} \bibnamefont{Chun}},
  \bibinfo{author}{\bibfnamefont{M.}~\bibnamefont{Passera}},
  \bibinfo{author}{\bibfnamefont{K.~M.} \bibnamefont{Patel}}, \bibnamefont{and}
  \bibinfo{author}{\bibfnamefont{S.~K.} \bibnamefont{Vempati}},
  \bibinfo{journal}{JHEP} \textbf{\bibinfo{volume}{11}}, \bibinfo{pages}{058}
  (\bibinfo{year}{2014}), \eprint{1409.3199}.

\bibitem[{\citenamefont{Chun and Kim}(2016)}]{Chun:2016hzs}
\bibinfo{author}{\bibfnamefont{E.~J.} \bibnamefont{Chun}} \bibnamefont{and}
  \bibinfo{author}{\bibfnamefont{J.}~\bibnamefont{Kim}},
  \bibinfo{journal}{JHEP} \textbf{\bibinfo{volume}{07}}, \bibinfo{pages}{110}
  (\bibinfo{year}{2016}), \eprint{1605.06298}.

\bibitem[{\citenamefont{Wang and Han}(2015)}]{Wang:2014sda}
\bibinfo{author}{\bibfnamefont{L.}~\bibnamefont{Wang}} \bibnamefont{and}
  \bibinfo{author}{\bibfnamefont{X.-F.} \bibnamefont{Han}},
  \bibinfo{journal}{JHEP} \textbf{\bibinfo{volume}{05}}, \bibinfo{pages}{039}
  (\bibinfo{year}{2015}), \eprint{1412.4874}.

\bibitem[{\citenamefont{Abe et~al.}(2015)\citenamefont{Abe, Sato, and
  Yagyu}}]{Abe:2015oca}
\bibinfo{author}{\bibfnamefont{T.}~\bibnamefont{Abe}},
  \bibinfo{author}{\bibfnamefont{R.}~\bibnamefont{Sato}}, \bibnamefont{and}
  \bibinfo{author}{\bibfnamefont{K.}~\bibnamefont{Yagyu}},
  \bibinfo{journal}{JHEP} \textbf{\bibinfo{volume}{07}}, \bibinfo{pages}{064}
  (\bibinfo{year}{2015}), \eprint{1504.07059}.

\bibitem[{\citenamefont{Crivellin et~al.}(2016)\citenamefont{Crivellin, Heeck,
  and Stoffer}}]{Crivellin:2015hha}
\bibinfo{author}{\bibfnamefont{A.}~\bibnamefont{Crivellin}},
  \bibinfo{author}{\bibfnamefont{J.}~\bibnamefont{Heeck}}, \bibnamefont{and}
  \bibinfo{author}{\bibfnamefont{P.}~\bibnamefont{Stoffer}},
  \bibinfo{journal}{Phys. Rev. Lett.} \textbf{\bibinfo{volume}{116}},
  \bibinfo{pages}{081801} (\bibinfo{year}{2016}), \eprint{1507.07567}.

\bibitem[{\citenamefont{Chun et~al.}(2015)\citenamefont{Chun, Kang, Takeuchi,
  and Tsai}}]{Chun:2015hsa}
\bibinfo{author}{\bibfnamefont{E.~J.} \bibnamefont{Chun}},
  \bibinfo{author}{\bibfnamefont{Z.}~\bibnamefont{Kang}},
  \bibinfo{author}{\bibfnamefont{M.}~\bibnamefont{Takeuchi}}, \bibnamefont{and}
  \bibinfo{author}{\bibfnamefont{Y.-L.~S.} \bibnamefont{Tsai}},
  \bibinfo{journal}{JHEP} \textbf{\bibinfo{volume}{11}}, \bibinfo{pages}{099}
  (\bibinfo{year}{2015}), \eprint{1507.08067}.

\bibitem[{\citenamefont{Han et~al.}(2016)\citenamefont{Han, Kang, and
  Sayre}}]{Han:2015yys}
\bibinfo{author}{\bibfnamefont{T.}~\bibnamefont{Han}},
  \bibinfo{author}{\bibfnamefont{S.~K.} \bibnamefont{Kang}}, \bibnamefont{and}
  \bibinfo{author}{\bibfnamefont{J.}~\bibnamefont{Sayre}},
  \bibinfo{journal}{JHEP} \textbf{\bibinfo{volume}{02}}, \bibinfo{pages}{097}
  (\bibinfo{year}{2016}), \eprint{1511.05162}.

\bibitem[{\citenamefont{Ilisie}(2015)}]{Ilisie:2015tra}
\bibinfo{author}{\bibfnamefont{V.}~\bibnamefont{Ilisie}},
  \bibinfo{journal}{JHEP} \textbf{\bibinfo{volume}{04}}, \bibinfo{pages}{077}
  (\bibinfo{year}{2015}), \eprint{1502.04199}.

\bibitem[{\citenamefont{Cherchiglia et~al.}(2018)\citenamefont{Cherchiglia,
  St\"ockinger, and St\"ockinger-Kim}}]{Cherchiglia:2017uwv}
\bibinfo{author}{\bibfnamefont{A.}~\bibnamefont{Cherchiglia}},
  \bibinfo{author}{\bibfnamefont{D.}~\bibnamefont{St\"ockinger}},
  \bibnamefont{and}
  \bibinfo{author}{\bibfnamefont{H.}~\bibnamefont{St\"ockinger-Kim}},
  \bibinfo{journal}{Phys. Rev. D} \textbf{\bibinfo{volume}{98}},
  \bibinfo{pages}{035001} (\bibinfo{year}{2018}), \eprint{1711.11567}.

\bibitem[{\citenamefont{Okada et~al.}(2016{\natexlab{b}})\citenamefont{Okada,
  Okada, Orikasa, and Yagyu}}]{Okada:2016whh}
\bibinfo{author}{\bibfnamefont{H.}~\bibnamefont{Okada}},
  \bibinfo{author}{\bibfnamefont{N.}~\bibnamefont{Okada}},
  \bibinfo{author}{\bibfnamefont{Y.}~\bibnamefont{Orikasa}}, \bibnamefont{and}
  \bibinfo{author}{\bibfnamefont{K.}~\bibnamefont{Yagyu}},
  \bibinfo{journal}{Phys. Rev. D} \textbf{\bibinfo{volume}{94}},
  \bibinfo{pages}{015002} (\bibinfo{year}{2016}{\natexlab{b}}),
  \eprint{1604.01948}.

\bibitem[{\citenamefont{Fan and Yagyu}(2022)}]{Fan:2022dye}
\bibinfo{author}{\bibfnamefont{Z.}~\bibnamefont{Fan}} \bibnamefont{and}
  \bibinfo{author}{\bibfnamefont{K.}~\bibnamefont{Yagyu}},
  \bibinfo{journal}{JHEP} \textbf{\bibinfo{volume}{06}}, \bibinfo{pages}{014}
  (\bibinfo{year}{2022}), \eprint{2201.11277}.

\bibitem[{\citenamefont{Aaltonen et~al.}(2022)}]{CDF:2022hxs}
\bibinfo{author}{\bibfnamefont{T.}~\bibnamefont{Aaltonen}} \bibnamefont{et~al.}
  (\bibinfo{collaboration}{CDF}), \bibinfo{journal}{Science}
  \textbf{\bibinfo{volume}{376}}, \bibinfo{pages}{170} (\bibinfo{year}{2022}).

\bibitem[{\citenamefont{Costantini et~al.}(2020)\citenamefont{Costantini,
  Ghezzi, and Pruna}}]{Costantini:2020xrn}
\bibinfo{author}{\bibfnamefont{A.}~\bibnamefont{Costantini}},
  \bibinfo{author}{\bibfnamefont{M.}~\bibnamefont{Ghezzi}}, \bibnamefont{and}
  \bibinfo{author}{\bibfnamefont{G.~M.} \bibnamefont{Pruna}},
  \bibinfo{journal}{Phys. Lett. B} \textbf{\bibinfo{volume}{808}},
  \bibinfo{pages}{135638} (\bibinfo{year}{2020}), \eprint{2001.08550}.

\bibitem[{\citenamefont{Gunion and Haber}(2003)}]{Gunion:2002zf}
\bibinfo{author}{\bibfnamefont{J.~F.} \bibnamefont{Gunion}} \bibnamefont{and}
  \bibinfo{author}{\bibfnamefont{H.~E.} \bibnamefont{Haber}},
  \bibinfo{journal}{Phys. Rev. D} \textbf{\bibinfo{volume}{67}},
  \bibinfo{pages}{075019} (\bibinfo{year}{2003}), \eprint{hep-ph/0207010}.

\bibitem[{\citenamefont{Branco et~al.}(2012)\citenamefont{Branco, Ferreira,
  Lavoura, Rebelo, Sher, and Silva}}]{Branco:2011iw}
\bibinfo{author}{\bibfnamefont{G.~C.} \bibnamefont{Branco}},
  \bibinfo{author}{\bibfnamefont{P.~M.} \bibnamefont{Ferreira}},
  \bibinfo{author}{\bibfnamefont{L.}~\bibnamefont{Lavoura}},
  \bibinfo{author}{\bibfnamefont{M.~N.} \bibnamefont{Rebelo}},
  \bibinfo{author}{\bibfnamefont{M.}~\bibnamefont{Sher}}, \bibnamefont{and}
  \bibinfo{author}{\bibfnamefont{J.~P.} \bibnamefont{Silva}},
  \bibinfo{journal}{Phys. Rept.} \textbf{\bibinfo{volume}{516}},
  \bibinfo{pages}{1} (\bibinfo{year}{2012}), \eprint{1106.0034}.

\bibitem[{\citenamefont{Maniatis et~al.}(2006)\citenamefont{Maniatis, von
  Manteuffel, Nachtmann, and Nagel}}]{Maniatis:2006fs}
\bibinfo{author}{\bibfnamefont{M.}~\bibnamefont{Maniatis}},
  \bibinfo{author}{\bibfnamefont{A.}~\bibnamefont{von Manteuffel}},
  \bibinfo{author}{\bibfnamefont{O.}~\bibnamefont{Nachtmann}},
  \bibnamefont{and} \bibinfo{author}{\bibfnamefont{F.}~\bibnamefont{Nagel}},
  \bibinfo{journal}{Eur. Phys. J. C} \textbf{\bibinfo{volume}{48}},
  \bibinfo{pages}{805} (\bibinfo{year}{2006}), \eprint{hep-ph/0605184}.

\bibitem[{\citenamefont{Ginzburg and Ivanov}(2005)}]{Ginzburg:2005dt}
\bibinfo{author}{\bibfnamefont{I.~F.} \bibnamefont{Ginzburg}} \bibnamefont{and}
  \bibinfo{author}{\bibfnamefont{I.~P.} \bibnamefont{Ivanov}},
  \bibinfo{journal}{Phys. Rev. D} \textbf{\bibinfo{volume}{72}},
  \bibinfo{pages}{115010} (\bibinfo{year}{2005}), \eprint{hep-ph/0508020}.

\bibitem[{\citenamefont{S\'anchez-Vega
  et~al.}(2019)\citenamefont{S\'anchez-Vega, Gambini, and
  Alvarez-Salazar}}]{Sanchez-Vega:2018qje}
\bibinfo{author}{\bibfnamefont{B.~L.} \bibnamefont{S\'anchez-Vega}},
  \bibinfo{author}{\bibfnamefont{G.}~\bibnamefont{Gambini}}, \bibnamefont{and}
  \bibinfo{author}{\bibfnamefont{C.~E.} \bibnamefont{Alvarez-Salazar}},
  \bibinfo{journal}{Eur. Phys. J. C} \textbf{\bibinfo{volume}{79}},
  \bibinfo{pages}{299} (\bibinfo{year}{2019}), \eprint{1811.00585}.

\bibitem[{\citenamefont{Eriksson et~al.}(2010)\citenamefont{Eriksson, Rathsman,
  and Stal}}]{Eriksson:2009ws}
\bibinfo{author}{\bibfnamefont{D.}~\bibnamefont{Eriksson}},
  \bibinfo{author}{\bibfnamefont{J.}~\bibnamefont{Rathsman}}, \bibnamefont{and}
  \bibinfo{author}{\bibfnamefont{O.}~\bibnamefont{Stal}},
  \bibinfo{journal}{Comput. Phys. Commun.} \textbf{\bibinfo{volume}{181}},
  \bibinfo{pages}{189} (\bibinfo{year}{2010}), \eprint{0902.0851}.

\bibitem[{\citenamefont{Enomoto and Watanabe}(2016)}]{Enomoto:2015wbn}
\bibinfo{author}{\bibfnamefont{T.}~\bibnamefont{Enomoto}} \bibnamefont{and}
  \bibinfo{author}{\bibfnamefont{R.}~\bibnamefont{Watanabe}},
  \bibinfo{journal}{JHEP} \textbf{\bibinfo{volume}{05}}, \bibinfo{pages}{002}
  (\bibinfo{year}{2016}), \eprint{1511.05066}.

\bibitem[{\citenamefont{Arbey et~al.}(2018)\citenamefont{Arbey, Mahmoudi, Stal,
  and Stefaniak}}]{Arbey:2017gmh}
\bibinfo{author}{\bibfnamefont{A.}~\bibnamefont{Arbey}},
  \bibinfo{author}{\bibfnamefont{F.}~\bibnamefont{Mahmoudi}},
  \bibinfo{author}{\bibfnamefont{O.}~\bibnamefont{Stal}}, \bibnamefont{and}
  \bibinfo{author}{\bibfnamefont{T.}~\bibnamefont{Stefaniak}},
  \bibinfo{journal}{Eur. Phys. J. C} \textbf{\bibinfo{volume}{78}},
  \bibinfo{pages}{182} (\bibinfo{year}{2018}), \eprint{1706.07414}.

\bibitem[{\citenamefont{Zyla et~al.}(2020)}]{ParticleDataGroup:2020ssz}
\bibinfo{author}{\bibfnamefont{P.~A.} \bibnamefont{Zyla}} \bibnamefont{et~al.}
  (\bibinfo{collaboration}{Particle Data Group}), \bibinfo{journal}{PTEP}
  \textbf{\bibinfo{volume}{2020}}, \bibinfo{pages}{083C01}
  (\bibinfo{year}{2020}).

\bibitem[{\citenamefont{de~Blas et~al.}(2022)\citenamefont{de~Blas, Pierini,
  Reina, and Silvestrini}}]{deBlas:2022hdk}
\bibinfo{author}{\bibfnamefont{J.}~\bibnamefont{de~Blas}},
  \bibinfo{author}{\bibfnamefont{M.}~\bibnamefont{Pierini}},
  \bibinfo{author}{\bibfnamefont{L.}~\bibnamefont{Reina}}, \bibnamefont{and}
  \bibinfo{author}{\bibfnamefont{L.}~\bibnamefont{Silvestrini}},
  \bibinfo{journal}{Phys. Rev. Lett.} \textbf{\bibinfo{volume}{129}},
  \bibinfo{pages}{271801} (\bibinfo{year}{2022}), \eprint{2204.04204}.

\bibitem[{\citenamefont{Lu et~al.}(2022)\citenamefont{Lu, Wu, Wu, and
  Zhu}}]{Lu:2022bgw}
\bibinfo{author}{\bibfnamefont{C.-T.} \bibnamefont{Lu}},
  \bibinfo{author}{\bibfnamefont{L.}~\bibnamefont{Wu}},
  \bibinfo{author}{\bibfnamefont{Y.}~\bibnamefont{Wu}}, \bibnamefont{and}
  \bibinfo{author}{\bibfnamefont{B.}~\bibnamefont{Zhu}},
  \bibinfo{journal}{Phys. Rev. D} \textbf{\bibinfo{volume}{106}},
  \bibinfo{pages}{035034} (\bibinfo{year}{2022}), \eprint{2204.03796}.

\bibitem[{\citenamefont{Grimus et~al.}(2008)\citenamefont{Grimus, Lavoura,
  Ogreid, and Osland}}]{Grimus:2008nb}
\bibinfo{author}{\bibfnamefont{W.}~\bibnamefont{Grimus}},
  \bibinfo{author}{\bibfnamefont{L.}~\bibnamefont{Lavoura}},
  \bibinfo{author}{\bibfnamefont{O.~M.} \bibnamefont{Ogreid}},
  \bibnamefont{and} \bibinfo{author}{\bibfnamefont{P.}~\bibnamefont{Osland}},
  \bibinfo{journal}{Nucl. Phys. B} \textbf{\bibinfo{volume}{801}},
  \bibinfo{pages}{81} (\bibinfo{year}{2008}), \eprint{0802.4353}.

\bibitem[{\citenamefont{Kim}(2022)}]{Kim:2022xuo}
\bibinfo{author}{\bibfnamefont{J.}~\bibnamefont{Kim}}, \bibinfo{journal}{Phys.
  Lett. B} \textbf{\bibinfo{volume}{832}}, \bibinfo{pages}{137220}
  (\bibinfo{year}{2022}), \eprint{2205.01437}.

\bibitem[{\citenamefont{Botella et~al.}(2022)\citenamefont{Botella,
  Cornet-Gomez, Mir\'o, and Nebot}}]{Botella:2022rte}
\bibinfo{author}{\bibfnamefont{F.~J.} \bibnamefont{Botella}},
  \bibinfo{author}{\bibfnamefont{F.}~\bibnamefont{Cornet-Gomez}},
  \bibinfo{author}{\bibfnamefont{C.}~\bibnamefont{Mir\'o}}, \bibnamefont{and}
  \bibinfo{author}{\bibfnamefont{M.}~\bibnamefont{Nebot}},
  \bibinfo{journal}{Eur. Phys. J. C} \textbf{\bibinfo{volume}{82}},
  \bibinfo{pages}{915} (\bibinfo{year}{2022}), \eprint{2205.01115}.

\bibitem[{\citenamefont{Benbrik et~al.}(2022)\citenamefont{Benbrik, Boukidi,
  and Manaut}}]{Benbrik:2022dja}
\bibinfo{author}{\bibfnamefont{R.}~\bibnamefont{Benbrik}},
  \bibinfo{author}{\bibfnamefont{M.}~\bibnamefont{Boukidi}}, \bibnamefont{and}
  \bibinfo{author}{\bibfnamefont{B.}~\bibnamefont{Manaut}}
  (\bibinfo{year}{2022}), \eprint{2204.11755}.

\bibitem[{\citenamefont{Ghorbani and Ghorbani}(2022)}]{Ghorbani:2022vtv}
\bibinfo{author}{\bibfnamefont{K.}~\bibnamefont{Ghorbani}} \bibnamefont{and}
  \bibinfo{author}{\bibfnamefont{P.}~\bibnamefont{Ghorbani}},
  \bibinfo{journal}{Nucl. Phys. B} \textbf{\bibinfo{volume}{984}},
  \bibinfo{pages}{115980} (\bibinfo{year}{2022}), \eprint{2204.09001}.

\bibitem[{\citenamefont{Babu et~al.}(2022)\citenamefont{Babu, Jana, and
  K.}}]{Babu:2022pdn}
\bibinfo{author}{\bibfnamefont{K.~S.} \bibnamefont{Babu}},
  \bibinfo{author}{\bibfnamefont{S.}~\bibnamefont{Jana}}, \bibnamefont{and}
  \bibinfo{author}{\bibfnamefont{V.~P.} \bibnamefont{K.}},
  \bibinfo{journal}{Phys. Rev. Lett.} \textbf{\bibinfo{volume}{129}},
  \bibinfo{pages}{121803} (\bibinfo{year}{2022}), \eprint{2204.05303}.

\bibitem[{\citenamefont{Song et~al.}(2022)\citenamefont{Song, Su, and
  Zhang}}]{Song:2022xts}
\bibinfo{author}{\bibfnamefont{H.}~\bibnamefont{Song}},
  \bibinfo{author}{\bibfnamefont{W.}~\bibnamefont{Su}}, \bibnamefont{and}
  \bibinfo{author}{\bibfnamefont{M.}~\bibnamefont{Zhang}},
  \bibinfo{journal}{JHEP} \textbf{\bibinfo{volume}{10}}, \bibinfo{pages}{048}
  (\bibinfo{year}{2022}), \eprint{2204.05085}.

\bibitem[{\citenamefont{Porod}(2003)}]{Porod:2003um}
\bibinfo{author}{\bibfnamefont{W.}~\bibnamefont{Porod}},
  \bibinfo{journal}{Comput. Phys. Commun.} \textbf{\bibinfo{volume}{153}},
  \bibinfo{pages}{275} (\bibinfo{year}{2003}), \eprint{hep-ph/0301101}.

\bibitem[{\citenamefont{Porod and Staub}(2012)}]{Porod:2011nf}
\bibinfo{author}{\bibfnamefont{W.}~\bibnamefont{Porod}} \bibnamefont{and}
  \bibinfo{author}{\bibfnamefont{F.}~\bibnamefont{Staub}},
  \bibinfo{journal}{Comput. Phys. Commun.} \textbf{\bibinfo{volume}{183}},
  \bibinfo{pages}{2458} (\bibinfo{year}{2012}), \eprint{1104.1573}.

\bibitem[{\citenamefont{Liu and Ng}(1994)}]{Liu:1993fwa}
\bibinfo{author}{\bibfnamefont{J.~T.} \bibnamefont{Liu}} \bibnamefont{and}
  \bibinfo{author}{\bibfnamefont{D.}~\bibnamefont{Ng}}, \bibinfo{journal}{Z.
  Phys. C} \textbf{\bibinfo{volume}{62}}, \bibinfo{pages}{693}
  (\bibinfo{year}{1994}), \eprint{hep-ph/9302271}.

\bibitem[{\citenamefont{Bechtle
  et~al.}(2014{\natexlab{a}})\citenamefont{Bechtle, Brein, Heinemeyer,
  St\r{a}l, Stefaniak, Weiglein, and Williams}}]{Bechtle:2013wla}
\bibinfo{author}{\bibfnamefont{P.}~\bibnamefont{Bechtle}},
  \bibinfo{author}{\bibfnamefont{O.}~\bibnamefont{Brein}},
  \bibinfo{author}{\bibfnamefont{S.}~\bibnamefont{Heinemeyer}},
  \bibinfo{author}{\bibfnamefont{O.}~\bibnamefont{St\r{a}l}},
  \bibinfo{author}{\bibfnamefont{T.}~\bibnamefont{Stefaniak}},
  \bibinfo{author}{\bibfnamefont{G.}~\bibnamefont{Weiglein}}, \bibnamefont{and}
  \bibinfo{author}{\bibfnamefont{K.~E.} \bibnamefont{Williams}},
  \bibinfo{journal}{Eur. Phys. J. C} \textbf{\bibinfo{volume}{74}},
  \bibinfo{pages}{2693} (\bibinfo{year}{2014}{\natexlab{a}}),
  \eprint{1311.0055}.

\bibitem[{\citenamefont{Bechtle et~al.}(2020)\citenamefont{Bechtle, Dercks,
  Heinemeyer, Klingl, Stefaniak, Weiglein, and Wittbrodt}}]{Bechtle:2020pkv}
\bibinfo{author}{\bibfnamefont{P.}~\bibnamefont{Bechtle}},
  \bibinfo{author}{\bibfnamefont{D.}~\bibnamefont{Dercks}},
  \bibinfo{author}{\bibfnamefont{S.}~\bibnamefont{Heinemeyer}},
  \bibinfo{author}{\bibfnamefont{T.}~\bibnamefont{Klingl}},
  \bibinfo{author}{\bibfnamefont{T.}~\bibnamefont{Stefaniak}},
  \bibinfo{author}{\bibfnamefont{G.}~\bibnamefont{Weiglein}}, \bibnamefont{and}
  \bibinfo{author}{\bibfnamefont{J.}~\bibnamefont{Wittbrodt}},
  \bibinfo{journal}{Eur. Phys. J. C} \textbf{\bibinfo{volume}{80}},
  \bibinfo{pages}{1211} (\bibinfo{year}{2020}), \eprint{2006.06007}.

\bibitem[{\citenamefont{Bechtle
  et~al.}(2014{\natexlab{b}})\citenamefont{Bechtle, Heinemeyer, St\r{a}l,
  Stefaniak, and Weiglein}}]{Bechtle:2013xfa}
\bibinfo{author}{\bibfnamefont{P.}~\bibnamefont{Bechtle}},
  \bibinfo{author}{\bibfnamefont{S.}~\bibnamefont{Heinemeyer}},
  \bibinfo{author}{\bibfnamefont{O.}~\bibnamefont{St\r{a}l}},
  \bibinfo{author}{\bibfnamefont{T.}~\bibnamefont{Stefaniak}},
  \bibnamefont{and} \bibinfo{author}{\bibfnamefont{G.}~\bibnamefont{Weiglein}},
  \bibinfo{journal}{Eur. Phys. J. C} \textbf{\bibinfo{volume}{74}},
  \bibinfo{pages}{2711} (\bibinfo{year}{2014}{\natexlab{b}}),
  \eprint{1305.1933}.

\bibitem[{\citenamefont{Bechtle et~al.}(2021)\citenamefont{Bechtle, Heinemeyer,
  Klingl, Stefaniak, Weiglein, and Wittbrodt}}]{Bechtle:2020uwn}
\bibinfo{author}{\bibfnamefont{P.}~\bibnamefont{Bechtle}},
  \bibinfo{author}{\bibfnamefont{S.}~\bibnamefont{Heinemeyer}},
  \bibinfo{author}{\bibfnamefont{T.}~\bibnamefont{Klingl}},
  \bibinfo{author}{\bibfnamefont{T.}~\bibnamefont{Stefaniak}},
  \bibinfo{author}{\bibfnamefont{G.}~\bibnamefont{Weiglein}}, \bibnamefont{and}
  \bibinfo{author}{\bibfnamefont{J.}~\bibnamefont{Wittbrodt}},
  \bibinfo{journal}{Eur. Phys. J. C} \textbf{\bibinfo{volume}{81}},
  \bibinfo{pages}{145} (\bibinfo{year}{2021}), \eprint{2012.09197}.

\bibitem[{\citenamefont{Hung et~al.}(2019)\citenamefont{Hung, Hong, Phuong,
  Mai, and Hue}}]{Hung:2019jue}
\bibinfo{author}{\bibfnamefont{H.~T.} \bibnamefont{Hung}},
  \bibinfo{author}{\bibfnamefont{T.~T.} \bibnamefont{Hong}},
  \bibinfo{author}{\bibfnamefont{H.~H.} \bibnamefont{Phuong}},
  \bibinfo{author}{\bibfnamefont{H.~L.~T.} \bibnamefont{Mai}},
  \bibnamefont{and} \bibinfo{author}{\bibfnamefont{L.~T.} \bibnamefont{Hue}},
  \bibinfo{journal}{Phys. Rev. D} \textbf{\bibinfo{volume}{100}},
  \bibinfo{pages}{075014} (\bibinfo{year}{2019}), \eprint{1907.06735}.

\bibitem[{\citenamefont{Aad et~al.}(2020)}]{ATLAS:2019nkf}
\bibinfo{author}{\bibfnamefont{G.}~\bibnamefont{Aad}} \bibnamefont{et~al.}
  (\bibinfo{collaboration}{ATLAS}), \bibinfo{journal}{Phys. Rev. D}
  \textbf{\bibinfo{volume}{101}}, \bibinfo{pages}{012002}
  (\bibinfo{year}{2020}), \eprint{1909.02845}.

\bibitem[{\citenamefont{Heinemeyer
  et~al.}(2004{\natexlab{a}})\citenamefont{Heinemeyer, Stockinger, and
  Weiglein}}]{Heinemeyer:2004yq}
\bibinfo{author}{\bibfnamefont{S.}~\bibnamefont{Heinemeyer}},
  \bibinfo{author}{\bibfnamefont{D.}~\bibnamefont{Stockinger}},
  \bibnamefont{and} \bibinfo{author}{\bibfnamefont{G.}~\bibnamefont{Weiglein}},
  \bibinfo{journal}{Nucl. Phys. B} \textbf{\bibinfo{volume}{699}},
  \bibinfo{pages}{103} (\bibinfo{year}{2004}{\natexlab{a}}),
  \eprint{hep-ph/0405255}.

\bibitem[{\citenamefont{Heinemeyer
  et~al.}(2004{\natexlab{b}})\citenamefont{Heinemeyer, Stockinger, and
  Weiglein}}]{Heinemeyer:2003dq}
\bibinfo{author}{\bibfnamefont{S.}~\bibnamefont{Heinemeyer}},
  \bibinfo{author}{\bibfnamefont{D.}~\bibnamefont{Stockinger}},
  \bibnamefont{and} \bibinfo{author}{\bibfnamefont{G.}~\bibnamefont{Weiglein}},
  \bibinfo{journal}{Nucl. Phys. B} \textbf{\bibinfo{volume}{690}},
  \bibinfo{pages}{62} (\bibinfo{year}{2004}{\natexlab{b}}),
  \eprint{hep-ph/0312264}.

\bibitem[{\citenamefont{Cherchiglia et~al.}(2017)\citenamefont{Cherchiglia,
  Kneschke, St\"ockinger, and St\"ockinger-Kim}}]{Cherchiglia:2016eui}
\bibinfo{author}{\bibfnamefont{A.}~\bibnamefont{Cherchiglia}},
  \bibinfo{author}{\bibfnamefont{P.}~\bibnamefont{Kneschke}},
  \bibinfo{author}{\bibfnamefont{D.}~\bibnamefont{St\"ockinger}},
  \bibnamefont{and}
  \bibinfo{author}{\bibfnamefont{H.}~\bibnamefont{St\"ockinger-Kim}},
  \bibinfo{journal}{JHEP} \textbf{\bibinfo{volume}{01}}, \bibinfo{pages}{007}
  (\bibinfo{year}{2017}), \bibinfo{note}{[Erratum: JHEP 10, 242 (2021)]},
  \eprint{1607.06292}.

\bibitem[{\citenamefont{Athron et~al.}(2022)\citenamefont{Athron, Balazs,
  Cherchiglia, Jacob, St\"ockinger, St\"ockinger-Kim, and
  Voigt}}]{Athron:2021evk}
\bibinfo{author}{\bibfnamefont{P.}~\bibnamefont{Athron}},
  \bibinfo{author}{\bibfnamefont{C.}~\bibnamefont{Balazs}},
  \bibinfo{author}{\bibfnamefont{A.}~\bibnamefont{Cherchiglia}},
  \bibinfo{author}{\bibfnamefont{D.~H.~J.} \bibnamefont{Jacob}},
  \bibinfo{author}{\bibfnamefont{D.}~\bibnamefont{St\"ockinger}},
  \bibinfo{author}{\bibfnamefont{H.}~\bibnamefont{St\"ockinger-Kim}},
  \bibnamefont{and} \bibinfo{author}{\bibfnamefont{A.}~\bibnamefont{Voigt}},
  \bibinfo{journal}{Eur. Phys. J. C} \textbf{\bibinfo{volume}{82}},
  \bibinfo{pages}{229} (\bibinfo{year}{2022}), \eprint{2110.13238}.

\bibitem[{\citenamefont{Athron et~al.}(2016)\citenamefont{Athron, Bach,
  Fargnoli, Gnendiger, Greifenhagen, Park, Pa\ss{}ehr, St\"ockinger,
  St\"ockinger-Kim, and Voigt}}]{Athron:2015rva}
\bibinfo{author}{\bibfnamefont{P.}~\bibnamefont{Athron}},
  \bibinfo{author}{\bibfnamefont{M.}~\bibnamefont{Bach}},
  \bibinfo{author}{\bibfnamefont{H.~G.} \bibnamefont{Fargnoli}},
  \bibinfo{author}{\bibfnamefont{C.}~\bibnamefont{Gnendiger}},
  \bibinfo{author}{\bibfnamefont{R.}~\bibnamefont{Greifenhagen}},
  \bibinfo{author}{\bibfnamefont{J.-h.} \bibnamefont{Park}},
  \bibinfo{author}{\bibfnamefont{S.}~\bibnamefont{Pa\ss{}ehr}},
  \bibinfo{author}{\bibfnamefont{D.}~\bibnamefont{St\"ockinger}},
  \bibinfo{author}{\bibfnamefont{H.}~\bibnamefont{St\"ockinger-Kim}},
  \bibnamefont{and} \bibinfo{author}{\bibfnamefont{A.}~\bibnamefont{Voigt}},
  \bibinfo{journal}{Eur. Phys. J. C} \textbf{\bibinfo{volume}{76}},
  \bibinfo{pages}{62} (\bibinfo{year}{2016}), \eprint{1510.08071}.

\bibitem[{\citenamefont{Aoyama et~al.}(2020)}]{Aoyama:2020ynm}
\bibinfo{author}{\bibfnamefont{T.}~\bibnamefont{Aoyama}} \bibnamefont{et~al.},
  \bibinfo{journal}{Phys. Rept.} \textbf{\bibinfo{volume}{887}},
  \bibinfo{pages}{1} (\bibinfo{year}{2020}), \eprint{2006.04822}.

\bibitem[{\citenamefont{Chakrabarty et~al.}(2014)\citenamefont{Chakrabarty,
  Dey, and Mukhopadhyaya}}]{Chakrabarty:2014aya}
\bibinfo{author}{\bibfnamefont{N.}~\bibnamefont{Chakrabarty}},
  \bibinfo{author}{\bibfnamefont{U.~K.} \bibnamefont{Dey}}, \bibnamefont{and}
  \bibinfo{author}{\bibfnamefont{B.}~\bibnamefont{Mukhopadhyaya}},
  \bibinfo{journal}{JHEP} \textbf{\bibinfo{volume}{12}}, \bibinfo{pages}{166}
  (\bibinfo{year}{2014}), \eprint{1407.2145}.

\bibitem[{\citenamefont{Chowdhury and Eberhardt}(2015)}]{Chowdhury:2015yja}
\bibinfo{author}{\bibfnamefont{D.}~\bibnamefont{Chowdhury}} \bibnamefont{and}
  \bibinfo{author}{\bibfnamefont{O.}~\bibnamefont{Eberhardt}},
  \bibinfo{journal}{JHEP} \textbf{\bibinfo{volume}{11}}, \bibinfo{pages}{052}
  (\bibinfo{year}{2015}), \eprint{1503.08216}.

\bibitem[{\citenamefont{Staub}(2014)}]{Staub:2013tta}
\bibinfo{author}{\bibfnamefont{F.}~\bibnamefont{Staub}},
  \bibinfo{journal}{Comput. Phys. Commun.} \textbf{\bibinfo{volume}{185}},
  \bibinfo{pages}{1773} (\bibinfo{year}{2014}), \eprint{1309.7223}.

\bibitem[{\citenamefont{Buttazzo et~al.}(2013)\citenamefont{Buttazzo, Degrassi,
  Giardino, Giudice, Sala, Salvio, and Strumia}}]{Buttazzo:2013uya}
\bibinfo{author}{\bibfnamefont{D.}~\bibnamefont{Buttazzo}},
  \bibinfo{author}{\bibfnamefont{G.}~\bibnamefont{Degrassi}},
  \bibinfo{author}{\bibfnamefont{P.~P.} \bibnamefont{Giardino}},
  \bibinfo{author}{\bibfnamefont{G.~F.} \bibnamefont{Giudice}},
  \bibinfo{author}{\bibfnamefont{F.}~\bibnamefont{Sala}},
  \bibinfo{author}{\bibfnamefont{A.}~\bibnamefont{Salvio}}, \bibnamefont{and}
  \bibinfo{author}{\bibfnamefont{A.}~\bibnamefont{Strumia}},
  \bibinfo{journal}{JHEP} \textbf{\bibinfo{volume}{12}}, \bibinfo{pages}{089}
  (\bibinfo{year}{2013}), \eprint{1307.3536}.

\end{thebibliography}

\end{document}